\documentclass[twocolumn, final]{elsarticle}

\usepackage{amssymb}
\usepackage{xcolor}
\usepackage{subcaption}
\usepackage{booktabs}

\biboptions{sort&compress}

\newpageafter{abstract}

\journal{Computational Materials Science}

\begin{document}

\begin{frontmatter}




\title{Lattice thermal conductivity of half-Heuslers with density functional theory and machine learning: Enhancing predictivity 
by active sampling with principal component analysis}


\author[NMBU_main]{Rasmus Tranås}

\affiliation[NMBU_main]{organization={Department of Mechanical Engineering and Technology Management, Norwegian University of Life Sciences},
            addressline={Drøbakveien 31}, 
            city={Ås},
            postcode={1432}, 
            state={Ås},
            country={Norway}}

\author[UiO,SINTEF]{Ole Martin Løvvik}

\affiliation[UiO]{organization={Centre for Materials Science and Nanotechnology, Department of Physics, University of Oslo},
            addressline={Sem Sælands vei 26}, 
            city={Oslo},
            postcode={0371}, 
            state={Oslo},
            country={Norway}}
            
\affiliation[SINTEF]{organization={Materials Physics, SINTEF},
            addressline={Forskningsveien 1}, 
            city={Oslo},
            postcode={0373}, 
            state={Oslo},
            country={Norway}}

\author[NMBU_additional]{Oliver Tomic}

\affiliation[NMBU_additional]{organization={Department of Data Science, Norwegian University of Life Sciences},
            addressline={Drøbakveien 31}, 
            city={Ås},
            postcode={1432}, 
            state={Ås},
            country={Norway}}

\author[NMBU_main]{Kristian Berland}

\begin{abstract}
Low lattice thermal conductivity is essential for 
high thermoelectric performance of a material. 
Lattice thermal conductivity is often computed using density functional theory (DFT),
typically at a high computational cost. 
Training machine learning models to predict lattice thermal conductivity could offer an effective procedure to identify low lattice thermal conductivity compounds.
However, in doing so, we must face the fact that such compounds can be quite rare and distinct from those in a typical training set.
This distinctness can be problematic as standard machine learning methods are inaccurate when predicting properties of compounds with features differing significantly from those in the training set. 
By computing the lattice thermal conductivity of 122 half-Heusler compounds, using the temperature-dependent effective potential method, we generate a data set to explore this issue. We first show how random forest regression can fail to identify low lattice thermal conductivity compounds with random selection of training data. Next, we show how active selection of training data
using feature and principal component analysis can be used to
improve model performance and the ability to identify low lattice thermal conductivity compounds. Lastly, we find that active learning without the use of DFT-based features can be viable as a quicker way of selecting samples.  
\end{abstract}

\end{frontmatter}

\clearpage
\newpage

\section{Introduction}
\label{sec:introduction}
With their ability to convert heat to electricity, thermoelectrics find use in several
niche technologies ranging from wine coolers, hiking stoves with mobile phone chargers, and radioisotope 
thermoelectric (TE) generators used to power e.g. the \textit{Curiosity} Mars rover. 
Thermoelectrics could also contribute to reducing global greenhouse gas emissions through waste heat recovery, but their role is currently limited by the modest efficacy realized in devices~\cite{snyderComplexThermoelectricMaterials2008, champierThermoelectricGeneratorsReview2017}.
Another limitation is the fact that several state-of-the-art TE materials contain toxic or rare elements~\cite{morenoReviewRecentProgress2020,weiReviewCurrentHighZT2020}.
Finding new TE materials has therefore gathered much scientific interest in recent years~\cite{recatala-gomezAcceleratedThermoelectricMaterials2020}.

The efficiency of TE materials is conventionally given by the dimensionless figure of merit, which is expressed as $ZT = \sigma S
^{2}T/(\kappa_{e} + \kappa_{\ell})$, where $\sigma$ is the electrical conductivity, $S$ is the Seebeck coefficient, $T$ is the absolute temperature, $\kappa_{e}$ is the electronic thermal conductivity, and $\kappa_\ell$ is the lattice thermal conductivity. High $ZT$ requires both a high power factor, $\mathcal{P} = \sigma S^{2}$, and low total thermal conductivity.
In non-metals, $\kappa_\ell$ is typically much larger than $\kappa_e$, but in heavily doped semiconductors,  $\kappa_\ell$ and $\kappa_e$ can be more comparable in size~\cite{paskovEffectSiDoping2017, kimInfluencePdDoping2019}; nonetheless, a low $\kappa_\ell$ is still typically needed for achieving high $ZT$.

High-throughput screening based on first-principle calculations have in recent years
been much used in the search for new TE materials~\cite{berlandThermoelectricTransportTrends2019,liHighThroughputScreeningAdvanced2019,choudharyDatadrivenDiscovery3D2020,liuHighthroughputDescriptorPrediction2020,heUltralowThermalConductivity2016,raghuvanshiHighThroughputSearch2020, JiaScreeningPromisingThermoelectric}. 
Many studies focus on electronic properties and use simple models or estimates of $\kappa_\ell$. One reason for this is that computing $\kappa_\ell$ comes at a significant computational cost.
The cost arises because accounting for the phonon-phonon interactions due to the anharmonicity of the lattice vibrations requires obtaining third-order force constants extracted from a large number of supercell-based density functional theory (DFT) calculations~\cite{togoDistributionsPhononLifetimes2015, hellmanTemperaturedependentEffectiveThirdorder2013, zhouLatticeAnharmonicityThermal2014}.
For this reason, machine learning (ML) methods are increasingly supplementing first-principles based calculations for predicting $\kappa_\ell$~\cite{carreteFindingUnprecedentedlyLowThermalConductivity2014, juExploringDiamondlikeLattice2019, chenMachineLearningModels2019, wangIdentificationCrystallineMaterials2020, loftisLatticeThermalConductivity2021, ZhuCharting2, visariaMachinelearningassistedSpacetransformationAccelerates2020, Gaultois2}. 
Pre-trained ML models can in turn also be made available in convenient web-based applications~\cite{gaultoisRecommendationEngineSuggesting2016}.

\begin{figure}[b!]
\centering
\includegraphics[width=0.75\linewidth]{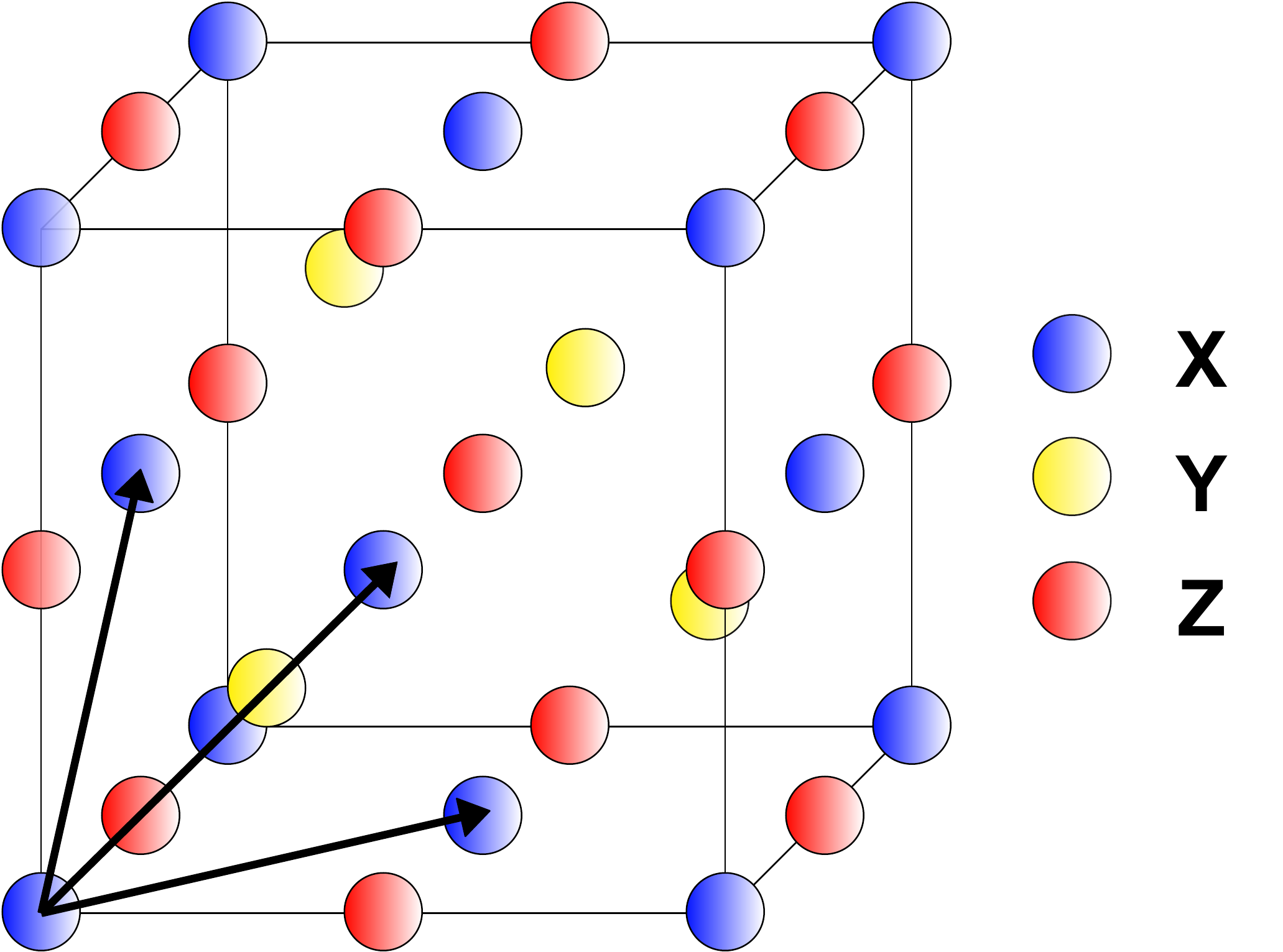}
\caption{\label{fig:HH_unit_cell}The HH crystal structure, displayed as the unit cell. The primitive cell is made from the three atoms $X$, $Y$, and $Z$.}
\end{figure}

The half-Heusler (HH) compounds are a class of cubic compounds with three atoms in the primitive cell, belonging to the 
$F\bar{4}3m$ spacegroup. As shown in Fig.~\ref{fig:HH_unit_cell}, the $XZ$ sublattice forms a rocksalt structure, while the $YZ$ sublattice forms a zinc-blende structure~\cite{casperHalfHeuslerCompoundsNovel2012,bosHalfHeuslerThermoelectricsComplex2014, zhuHighEfficiencyHalfHeusler2015}.
Several HH compounds have a high power factor in combination with relatively low $\kappa_\ell$, making HHs attractive for TE applications~\cite{berlandThermoelectricTransportTrends2019, yuanEffectsSbSubstitution2017, horiFirstprinciplesCalculationLattice2020, raghuvanshiHighThroughputSearch2020, chauhanDefectEngineeringEnhancement2020, zhuDiscoveryTaFeSbbasedHalfHeuslers2019, sunRemarkablyHighThermoelectric2020, zhaoSynthesisThermoelectricProperties2014, Zhou2018LargeTP}.
Recently, Feng et al.~\cite{fengCharacterizationRattlingRelation2020} used DFT-based calculations  
to show that the four HH compounds: CdPNa, BaBiK, LaRhTe, and LaPtSb have very low $\kappa_\ell$. LaPtSb and BaBiK also have promising electronic transport properties in addition to low $\kappa_\ell$ and could have a $ZT$ competitive with top performing TE materials~\cite{xueLaPtSbHalfHeuslerCompound2016,hanHighThermoelectricPerformance2020,samsonidzeAcceleratedScreeningThermoelectric2018}. 
Because $\kappa_\ell$ of these four compounds is much lower than for typical HHs, the set of HH compounds presents itself as a dataset well suited for investigating ML methods to separate low and high $\kappa_\ell$ compounds.
The high symmetry of HHs also reduces the computational cost of calculating $\kappa_\ell$ compared to more complex systems such as layered compounds and compounds with distorted symmetries~\cite{zhangFirstprinciplesStudyLayered2019, sarkarFerroelectricInstabilityInduced2020, zhaoUltralowThermalConductivity2014}. The reduced cost
allows us to generate both training and test sets for assessing $\kappa_\ell$. 

Our study is based on 122 HH compounds for which $\kappa_{\ell}$ is computed explicitly using DFT. The compounds are based on a combination of 
dynamically stable HHs, 54 from groups 4-9-15 (Ti,Zr,Hf)(Co,Rh,Ir)(As,Sb,Bi), 4-10-14
(Ti,Zr,Hf)(Ni,Pd,Pt)(Ge,Sn,Pb), and 48 HHs from groups 5-8-15 (V,Nb,Ta)(Fe,Ru,Os)(As,Sb,Bi) and 5-9-14 (V,Nb,Ta)(Co,Rh,Ir)(Ge,Sn,Pb). The last 20 HHs are the remaining stable compounds studied by 
Feng et al. \cite{fengCharacterizationRattlingRelation2020}, based on a revision of the 75 stable HHs identified by Carrete et al. \cite{carreteFindingUnprecedentedlyLowThermalConductivity2014}. 

\section{\label{sec:methods}Methods}

\subsection{\label{sec:ltc}Lattice thermal conductivity}

DFT calculations in this work are done with the \textsc{VASP} \cite{kresseInitioMolecularDynamics1993,kresseEfficiencyAbinitioTotal1996,kresseEfficientIterativeSchemes1996} software package using the Perdew-Burke-Ernzerhof (PBE) generalized gradient approximation for solids, PBEsol~\cite{perdewRestoringDensityGradientExpansion2008,csonkaAssessingPerformanceRecent2009}. 
The plane-wave energy cutoff is set to 600 eV.
For relaxations, we use an $11\times 11\times 11$ $\mathbf{k}$-point sampling of the Brillouin zone. 
The electronic self-consistent loop is iterated until the energy difference falls below $10^{-6}$~eV, while ionic positions are relaxed until forces fall below 1~$\mathrm{meV}/\mathrm{Å}$. 
The lattice thermal conductivity, $\kappa_\ell$, is calculated with the temperature-dependent effective potential (TDEP) method~\cite{hellmanLatticeDynamicsAnharmonic2011,hellmanTemperaturedependentEffectiveThirdorder2013}, taking into account three-phonon and isotope-phonon scattering \cite{tamuraIsotopeScatteringDispersive1983, tamuraIsotopeScatteringLargewavevector1984}.
Fifty configurations based on $3 \times 3 \times 3$ repetitions of the primitive cell are used to obtain second- and third-order force constants. 
The atomic configurations are taken from a fixed-temperature canonical ensemble at $300$~K, where the zero-point motion of the phonons is matched with the Debye temperature~\cite{shulumbaIntrinsicLocalized}. 
The Debye temperature is obtained from the Voigt approximation of the bulk and shear moduli~\cite{AndersonSimplifiedMethod}. 
A $3\times3\times3$ $\mathbf{k}$-point grid is used for the supercell DFT force calculations. 
We employ a cutoff for second-order pair-interactions of 7~\AA\, while for third-order pair-interactions, the cutoff is set slightly larger than half the width of the supercell (i.e. 6.1~\AA\, for NbCoGe). For the calculation of $\kappa_\ell$, the reciprocal space is discretized on a $35\times35\times35$ $\mathbf{q}$-point grid. 
In a convergence study for NbCoGe, we find these cutoffs to give a numerical error of $\kappa_\ell$ less than 3 \%. 

\subsection{\label{sec:rfr} Machine learning model }

Random forest (RF) regression is a non-linear ML method used in industry and academia alike~\cite{breiman_rf}.
An ensemble of decision trees forms the RF model, where each tree is trained on a subset of randomly chosen features and training samples. This randomness makes RF less prone to overfitting.
RF has been shown to perform well in earlier ML studies involving the lattice thermal conductivity~\cite{ZhuCharting2}. In the RF regression, a given input sample 
is sorted in each of the decision trees based on its features, 
so that in a given tree, the sample is assigned to a $\kappa_\ell^{\{i..\}}$ 
in the training set. Finally, the predicted outcome is given by the mean $\langle \kappa_\ell^{\{i..\}} \rangle$ of the predictions of the ensemble of decision trees. 

In ML, failing to identify key features can result in overfitting and reduce method interpretability~\cite{python_ml, jabbarMethodsAvoidOverFitting2014}.
Feature selection is here performed using exhaustive feature selection (EFS) in combination with RF regression.
EFS assesses the predictive performance of every subset of extracted features and finds the features that give the best outcome of a chosen performance metric. We here choose to use Spearman rank correlation as the metric with five-fold cross-validation, as this correlation measures the predicted ranking of compounds.
This brute-force approach carries a significant computational cost, 
but with the limited number of features in our study, this cost is small compared to that of computing $\kappa_\ell$. 
EFS is done with the \textsc{MLxtend}~\cite{raschkaMLxtendProvidingMachine2018} code, while RF regression is done using
\textsc{Scikit-learn}~\cite{pedregosaScikitlearnMachineLearning2011}. 
In the RF model, the hyperparameters of each set of features are optimized using a hyperparameter grid search.

\begin{figure}[h!]
\includegraphics[width=\linewidth]{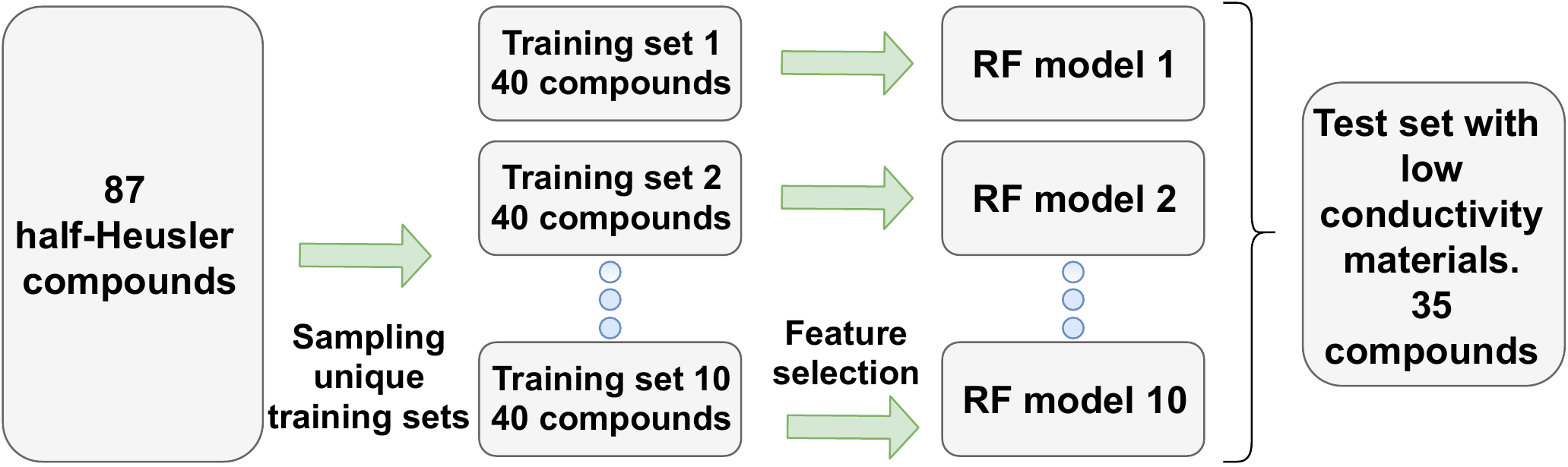}
\caption{\label{fig:ML_flow_chart}Flowchart for building the baseline model.}
\end{figure}

Fig.~\ref{fig:ML_flow_chart} shows a flowchart for the baseline model.
In the model, 87 of the 122 compounds are semi-randomly selected as the training pool for ML, while 35 are left out to provide a test set for model assessment. 
By semi-randomly, we refer to the fact the five lowest $\kappa_\ell$ compounds are in the test set. 
We make this choice to emulate a not too improbable scenario that could easily arise for larger material classes when only modest-size training sets are used. 
From the training pool, 10 unique training sets of 40 compounds are selected randomly.
The models are retrained based on the features obtained with EFS for final model evaluation. Our baseline model predictions are given by the average of the predictions of these 10 RF models. 

In the active sampling scheme, we 
use principal component analysis (PCA)
with the \textsc{hoggorm}~\cite{tomicHoggormPythonLibrary2019} package
to identify compounds possessing combinations of feature values that are distinct from those in the training pool.
PCA accounts for correlations between features by constructing orthogonal principal components (PC) as linear combinations of feature vectors in feature space. 
The PCs are oriented in the direction of maximum variance and the features
are centered and scaled to unit variance. The PCA analysis is based on all compounds in the study. 
Using PCA, we identify three compounds, BaBiK, CdPNa, and LaPtSb, that are needed to cover the feature space
mapped out by the first two PCs.
These three are subsequently included in the training sets from the baseline model, such that the 10 training sets for the active sampling model contain 43 compounds that are used with RF and EFS.

Our study is based on 14 features: 9 are tabulated while 5 are obtained from lost-cost DFT calculations. Two of these, 
the volume of the relaxed primitive unit cell, $V$, and corresponding mass density, $\rho$, could in principle have been obtained from typical tabulated data, standard experiment, or in the absence of such data, from ML models~\cite{miyazakiMachineLearningBased2021, LiangCRYSPNet, LiMlatticeabc}. The tabulated features together with $V$ and $\rho$ are grouped as the easily available tier-0 features. 
Tabulated features are as follows:
the ratio between the lightest and heaviest atoms in the primitive cell~\cite{MeiJaAtomicWeights}, $m_{r}$, the average atomic mass, $m_{a}$, the standard deviation of the atomic masses, $m_{s} = 1/3 \left( {\sum_{i=X,Y,Z}(m_{i} - m_{a})^2} \right)^{1/2 }$, 
as well as corresponding features for the electronegativity~\cite{AllenElectronegativity}, $\chi$, and covalent atomic radius~\cite{CorderoCovalentRadii}, $r$.
The remaining three, which together with the tier-0 features constitute the tier-1 features, are the lattice thermal conductivity in the Slack model~\cite{slackNonmetallicRystalsHigh1973}, $\kappa_{s}$, the Debye temperature, $\theta_{D}$, and the bulk modulus, $B$. 
These three are related to the elastic tensor~\cite{jiaLatticeThermalConductivity2017}, and are hence the most time-consuming features to generate.
Sections \ref{sec:pca_edge_point} through \ref{sec:active_sample_model} are based on tier-1 features, while Section \ref{sec:active_sample_model_without_DFT_features} compares the ML performance of models based on tier-0 and tier-1 features.


Higher-order features beyond what we consider, such as the three-phonon scattering phase space, effective spring constants, and first moment frequencies from the phonon density of states~\cite{carreteFindingUnprecedentedlyLowThermalConductivity2014,juMaterialsInformaticsHeat2019,fengCharacterizationRattlingRelation2020}, 
can improve the predictions of the ML model, but we here limit ourselves to features that are based on properties that one can expect to be continuously added in material databases such as the \textsc{MaterialsProject}~\cite{jainCommentaryMaterialsProject2013}. Therefore, using such simple features supports a methodology that can later be adopted for screening of larger material databases.  

\section{\label{sec:results_DFT}Results: Density functional theory calculations}

\begin{figure*}[t!]
   \begin{subfigure}[b]{\columnwidth}
        \centering
        \includegraphics[height=2.35in]{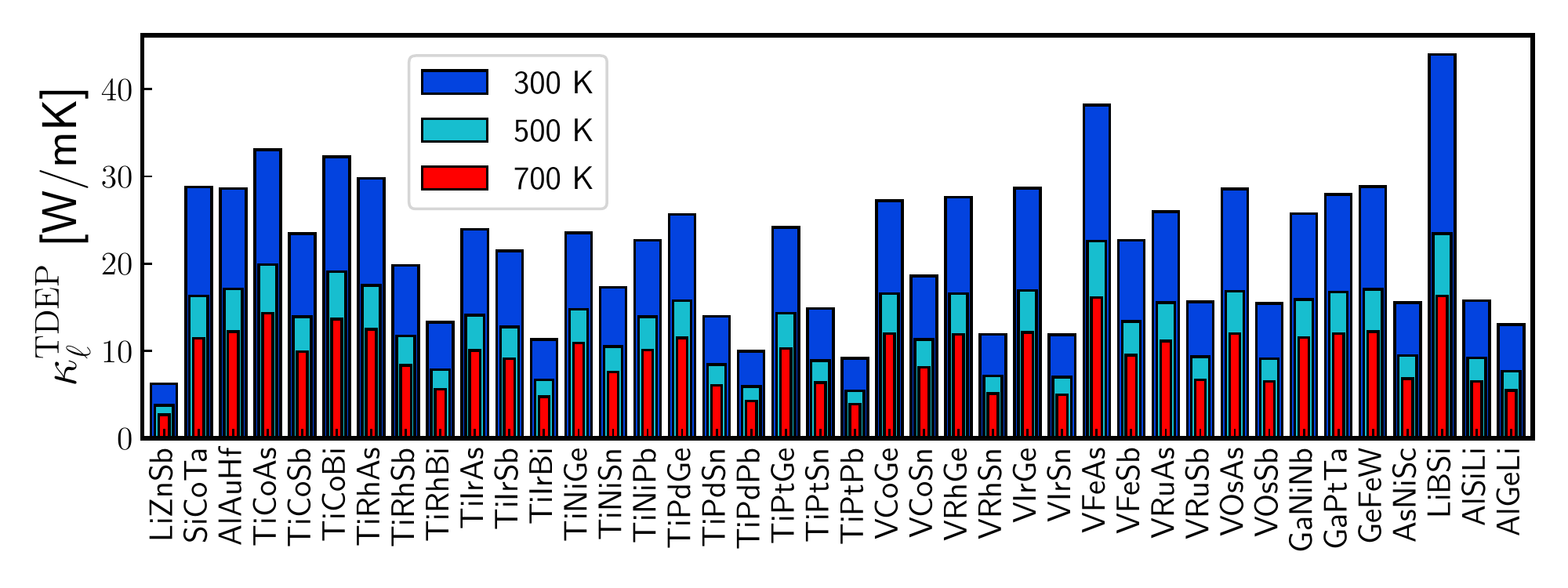}
        \caption*{\label{fig:TiV}}
    \end{subfigure}%
    \vspace{-2.1\baselineskip}
    \\
    \begin{subfigure}[b]{\columnwidth}
        \centering
        \includegraphics[height=2.35in]{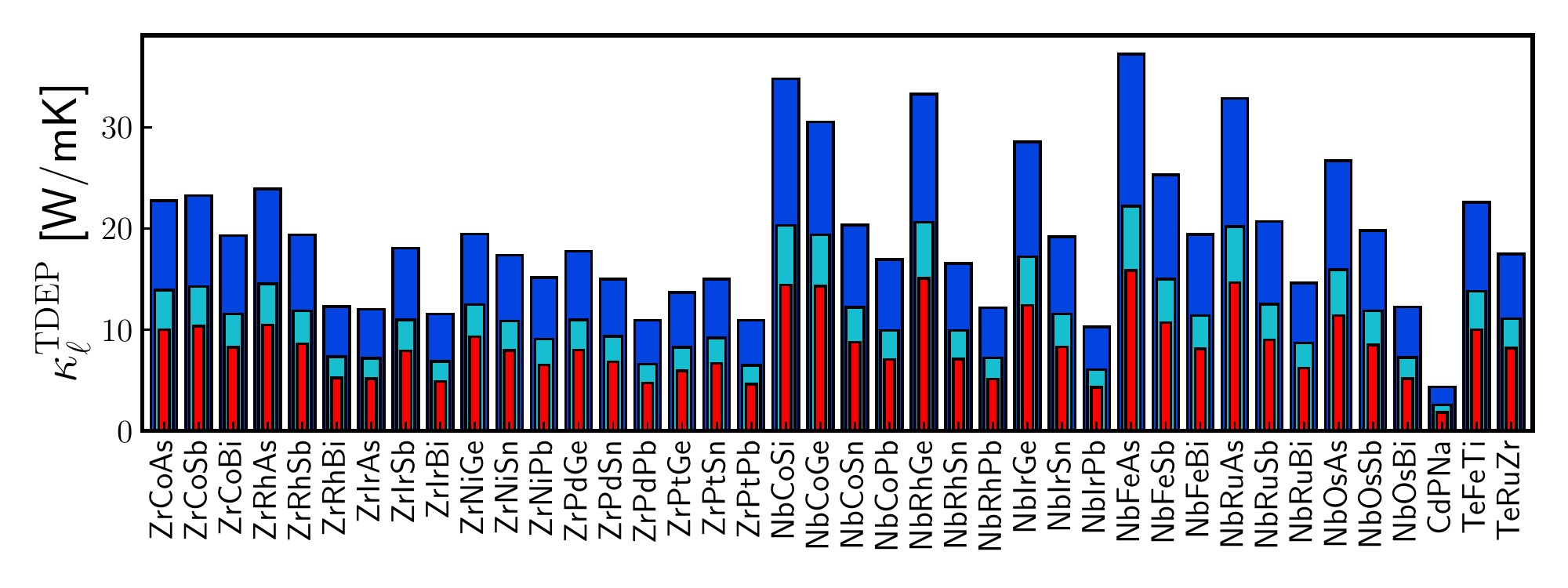}
        \caption*{\label{fig:ZrNb}}
    \end{subfigure}%
    \vspace{-2.1\baselineskip}
    \\
    \begin{subfigure}[b]{\columnwidth}

        \includegraphics[height=2.35in]{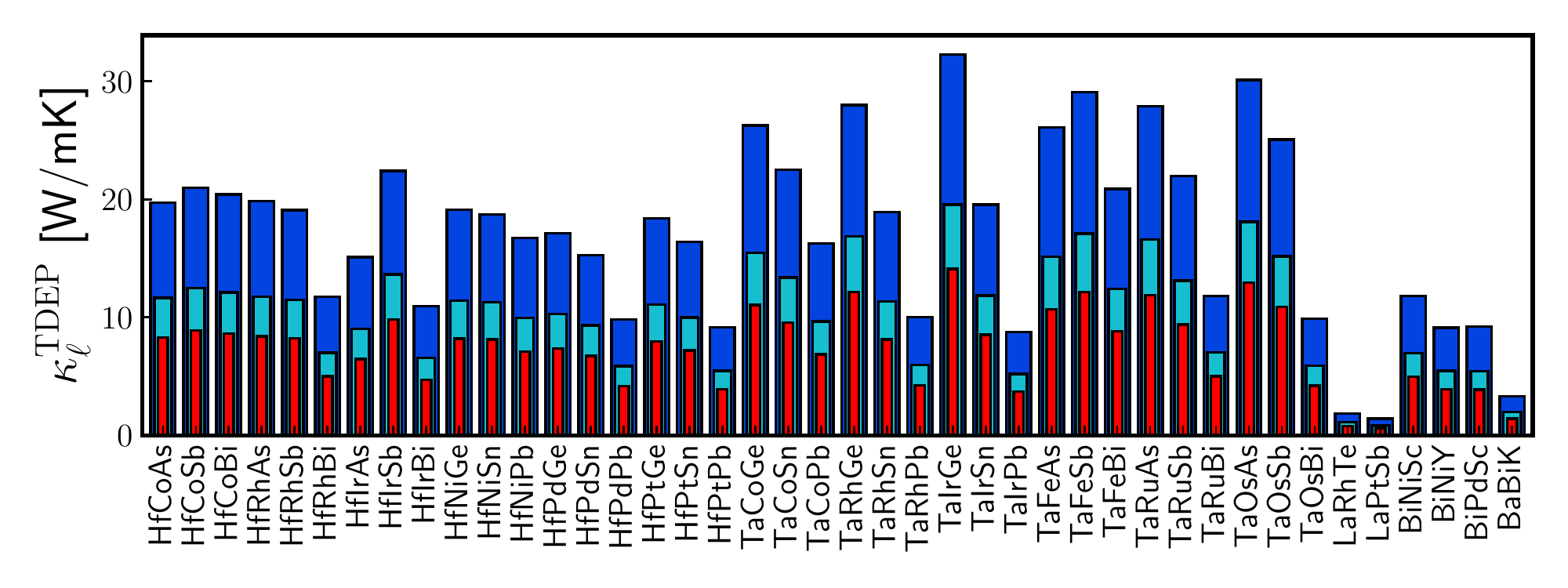}
        \caption*{\label{fig:TiV}}
    \end{subfigure}%
\caption{\label{fig:kappa_triplot}$\kappa_{\ell}^{\mathrm{TDEP}}$ for the HHs at 300 K (blue bars), 500 K (turquoise bars), and 700 K (red bars).}
\end{figure*}

\subsection{\label{sec:ltc_results}Lattice thermal conductivity of half-Heusler compounds}

Fig.~\ref{fig:kappa_triplot} shows the lattice thermal conductivity calculated with TDEP, $\kappa_{\ell}^{\mathrm{TDEP}}$, at 300~K, 500~K, and 700~K for the 122 HHs. \ref{app:A}: Table~\ref{table:kappas} reports $\kappa_{\ell}^{\mathrm{TDEP}}$ at 500~K. 
At 500~K, the span of $\kappa_{\ell}^{\mathrm{TDEP}}$ goes from 0.85~W/mK (LaPtSb) to 23.45~W/mK (LiBSi).
Compounds with heavy atoms on the $X$- or $Z$-site, such as La, Ba, Bi, and Pb, or with high average mass, typically have lower $\kappa_{\ell}^{\mathrm{TDEP}}$. The correlation between lattice thermal conductivity and the average mass has been observed for experimental lattice thermal conductivity with compounds across different spacegroups~\cite{chenMachineLearningModels2019}. The ordering of the five lowest $\kappa_{\ell}^{\mathrm{TDEP}}$ materials from low to high $\kappa_{\ell}^{\mathrm{TDEP}}$ is consistent with the findings of Feng et al.~\cite{fengCharacterizationRattlingRelation2020}. 
The Vanadium-containing compounds VRuBi, VFeBi, VIrPb, VOsBi, VRhPb, and VCoPb have negative phonon frequencies and are not studied further. The two latter have previously been predicted to decompose into elemental phases~\cite{gautierPredictionAcceleratedLaboratory2015}.

In the following, the ML models are based on $\kappa_{\ell}^{\mathrm{TDEP}}$ at 500~K, which also indicates low lattice thermal conductivity at 300~K and 700~K.

\section{\label{sec:results_DFT}Results and discussion: Machine learning}

\subsection{\label{sec:pca_edge_point}Using principal component analysis for diversifying training sets}
\begin{figure}[t!]
    \centering
    \begin{subfigure}[b]{\columnwidth}
        \centering
        \includegraphics[height=2.25in]{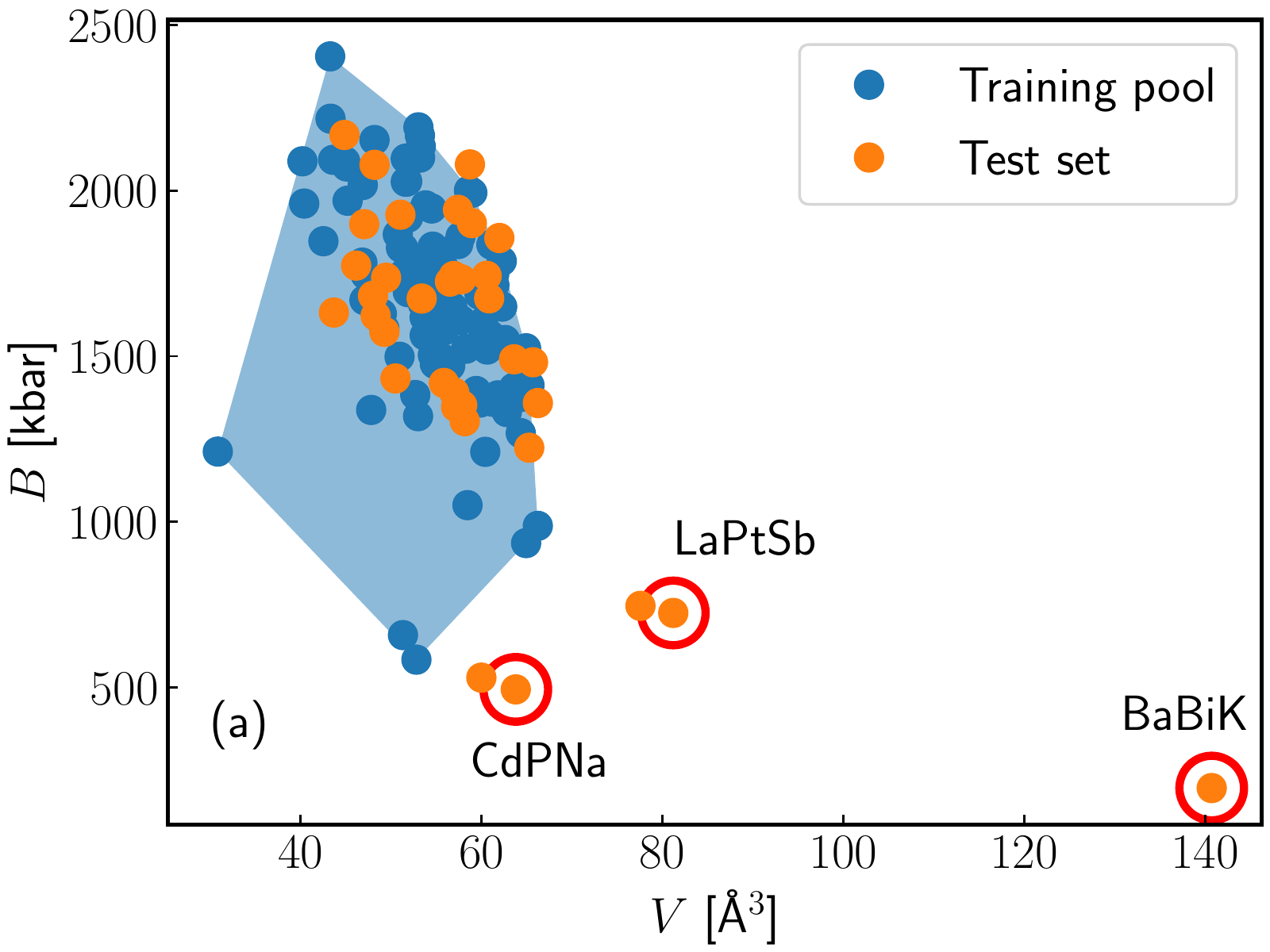}
        \caption*{\label{fig:feature_correlation}}
    \end{subfigure}%
    \\
    \begin{subfigure}[b]{\columnwidth}
        \centering
        \includegraphics[height=2.25in]{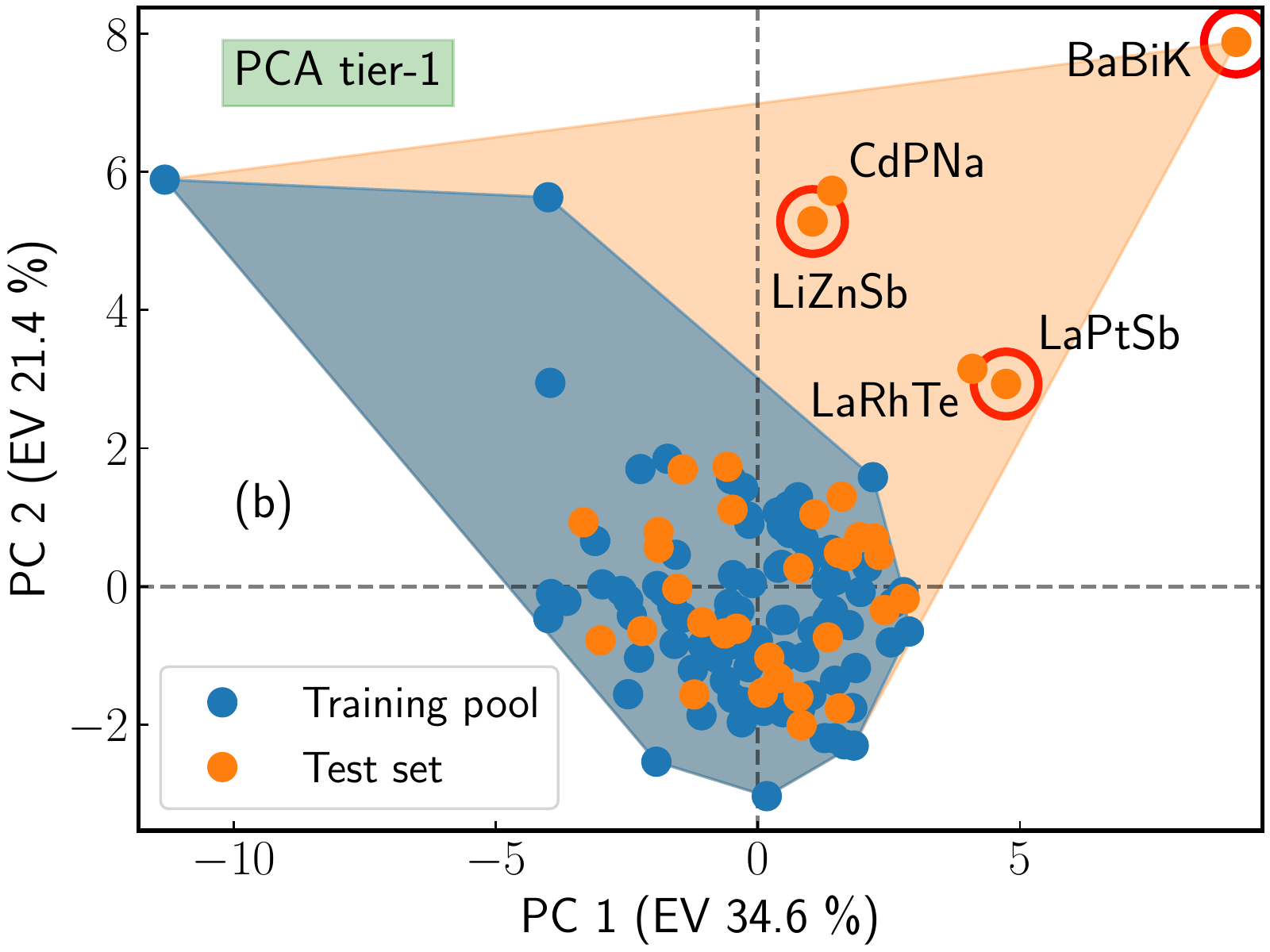}
        \caption*{\label{fig:pca_two_comp}}
    \end{subfigure}
    \caption{\label{fig:corr_and_pca}(a) Scatter plot for the compounds, with $V$ on the horizontal axis and $B$ on the vertical axis. (b) PCA plot for the compounds using tier-1 features. The first PC is shown on the horizontal axis and the second is on the vertical axis. The blue convex hull is the area of the PC space spanned by the 87 materials in the training pool, while the orange convex hull is the area spanned with inclusion of the test compounds.}
\end{figure}


Fig.~\ref{fig:corr_and_pca}~(a) shows the position of the compounds in the space spanned by $V$ and $B$, two features which are known to correlate with the lattice thermal conductivity
\cite{gaultois3, carreteFindingUnprecedentedlyLowThermalConductivity2014, chenMachineLearningModels2019}.
In general, low $V$ materials tend to have less compressed acoustical phonon band structures, increasing the phonon group velocity~\cite{anInitioPhononDispersions2008} and thus also the lattice thermal conductivity. 
Higher $B$ can also be related to stiffer atomic bonds and increased phonon velocities. 
This is reflected in a Spearman correlation of $-0.69$ between $\kappa_\ell^\mathrm{TDEP}$ and $V$
and 0.58 between $\kappa_\ell^\mathrm{TDEP}$ and $B$ for the compounds in the training pool.
The plot shows that some of the compounds in the test set fall outside the convex hull spanned by $V$ and $B$ of the compounds in the training pool. 
Including such outliers in the training sets, could result in more accurate ML predictions. 
However, as $B$ and $V$ also have a Spearman correlation of $-0.51$,
---i.e. higher $V$ tend to relate to less stiff bonds and thus lower $B$ ---
solely relying on these two features could risk missing important compounds and correlations.
This motivates the use of PCA, which offers a more systematic procedure to take all features and their correlations into consideration.

Fig.~\ref{fig:corr_and_pca}~(b) indicates the position of the compounds of the test set and training pool in the subspace spanned by the two first PCs. This subspace accounts for 56.0~\% of the cumulative explained variance (EV) of the feature space. Mapping these two features back to the original feature space, we find the cumulative explained variances of $V$ and $B$ to be 78.6~\% and 77.2~\%. 
A comparison of the two convex hulls shows that the low $\kappa_\ell$ compounds lie outside of convex hull spanned by the training pool.

While PCA  can support a human-guided selection of training sets, we choose to formalize this in a systematic procedure that is better suited for automation of the active sampling.
The specific compounds to be included are determined iteratively by identifying the compound in the test set with the largest Euclidean distance in the PC space to the closest compound in the current training pool until a marked drop in distance arises. The procedure identifies that three additional compounds, BaBiK, CdPNa, and LaPtSb, should be included in the training process. 
These compounds are highlighted with red circles in Fig.~\ref{fig:corr_and_pca}, and
Fig.~\ref{fig:PCA_distance} shows the distance in PC space after each iteration.

\subsection{\label{sec:feature_selection}Exhaustive feature selection analysis}

\begin{figure}[t!]
\includegraphics[width=\linewidth]{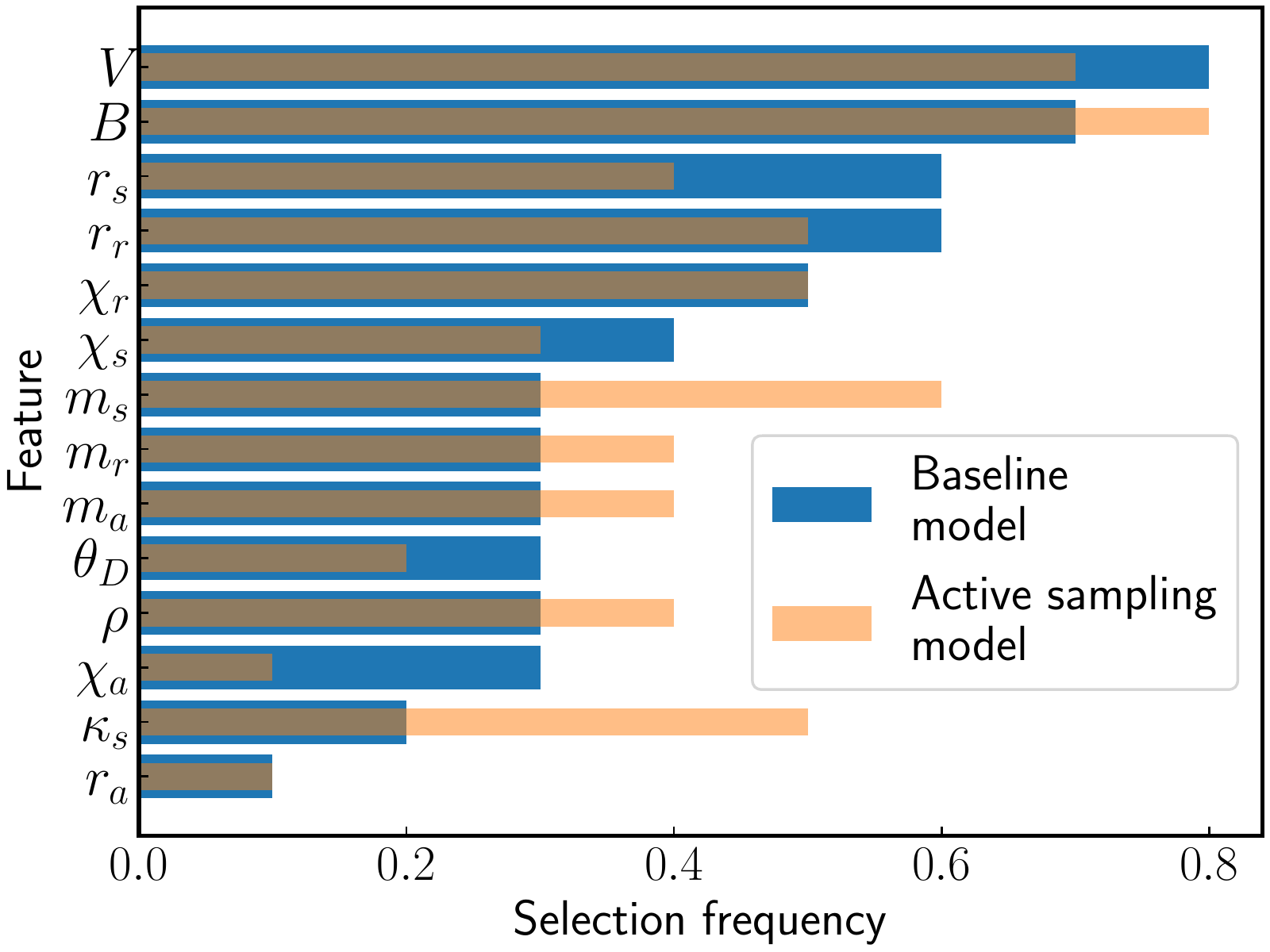}
\caption{\label{fig:feature_importance}
Selection frequencies for the 14 features in the EFS using RF regression for the baseline (blue bars) and active sampling (orange bars) models.}
\end{figure}

Fig.~\ref{fig:feature_importance} compares the EFS feature selection frequency of the baseline and active sampling ML models. 
The baseline and active sampling models use on average 5.7 and 5.9 features out of the 14 potential features, respectively. 
The relatively few features selected is in line with the recent results of Miyazaki et al.~\cite{miyazakiMachineLearningBased2021} 
finding that using a limited subset of features gives the best ML performance, which can be linked to the fact that redundant features can cause overfitting. 
In both the baseline and active sampling models, $B$ and $V$ are the most frequently selected 
features, in agreement with their high Spearman correlation with $\kappa_\ell^\mathrm{TDEP}$.

There are some notable differences between the EFS for the active sampling model and the baseline model. 
In particular, the selection frequency of $m_s$ increases from 0.3 to 0.6 for the active sampling model.
This result reflects that the variation of masses in the primitive cell
is linked to low lattice thermal conductivity, such as for BaBiK. The selection frequencies for $\kappa_{s}$ and $B$ also increase for the active sampling model.

\subsection{\label{sec:active_sample_model}Enhanced machine learning performance with active sampling}

\begin{figure}[h!]
\begin{subfigure}[b]{\columnwidth}
\includegraphics[width=\linewidth]{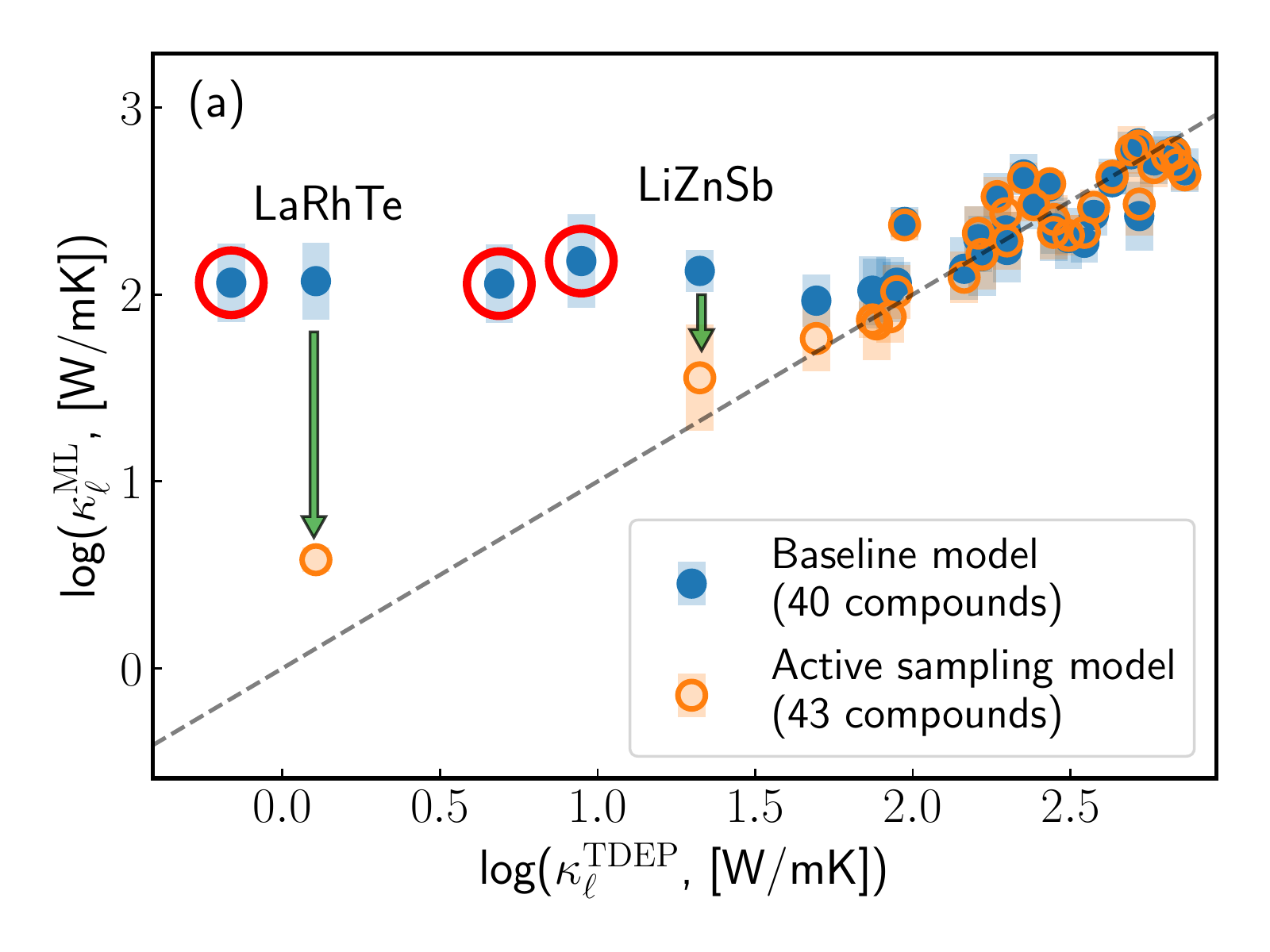}
\caption*{\label{fig:rfr_baseline}}
\end{subfigure}
\vspace{-2\baselineskip}
\\
\begin{subfigure}[b]{\columnwidth}
\includegraphics[width=\linewidth]{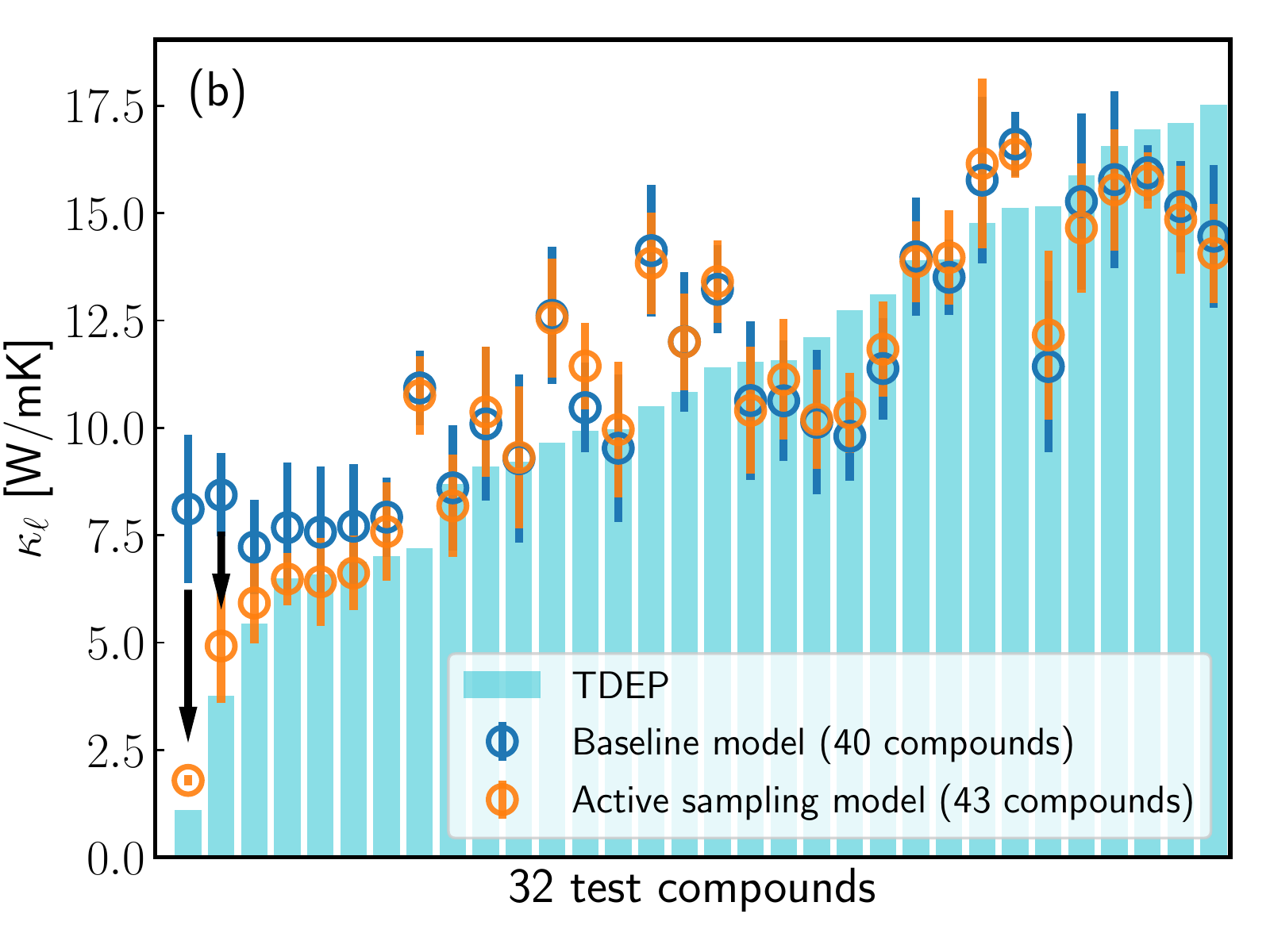}
\caption*{\label{fig:rfr_w_dualplot}}
\end{subfigure}
\vspace{-2\baselineskip}
\caption{
a) Parity plot for predictions made on the test set. The horizontal axis shows $\log(\kappa_\ell^{\mathrm{TDEP}})$ at 500 K, while the vertical axis shows the predictions. Blue (orange) circles indicate the predictions made by the baseline (active sampling) model. b) Corresponding comparison for $\kappa_\ell^{\mathrm{TDEP}}$.  The turquoise bars indicate $\kappa_\ell^{\mathrm{TDEP}}$ at 500 K.
}\label{fig:active_vs_baseline}
\end{figure}

Fig.~\ref{fig:active_vs_baseline}~(a) compares the predictions of the baseline model and the active sampling model on a logarithmic scale as used in the training. 
The error bars indicate the standard deviation of the predictions of the 10 models. The figure shows that the active sampling model has a superior ability to identify the compounds with low $\kappa_\ell^{\rm TDEP}$. Predictions for the three compounds found with PCA, highlighted with red circles, are not provided for the active sampling model as they are included in the training sets of the model.
Fig.~\ref{fig:active_vs_baseline}~(b) shows the corresponding comparison with a linear scale, with compounds sorted according to $\kappa_\ell^{\mathrm{TDEP}}$. In most of the cases, the active sampling model predictions, $\kappa_\ell^{\mathrm{AS}}$, are higher than $\kappa_\ell^{\mathrm{TDEP}}$ for low $\kappa_\ell^{\mathrm{TDEP}}$ compounds, and vice versa for high $\kappa_\ell^{\mathrm{TDEP}}$ compounds. 
This is also seen in the logarithmic scale of Fig.~\ref{fig:active_vs_baseline}~(a).
Even if the numerical precision of the active sampling model for the compounds with low $\kappa_\ell^{\mathrm{TDEP}}$ is quite modest, which can be linked to the limited sampling in this region of feature space,
the model identifies the compounds with the lowest $\kappa_\ell^{\rm TDEP}$. 
\ref{app:A}: Table~\ref{table:kappas} provides the numerical values of $\kappa_\ell^{\mathrm{TDEP}}$ and $\kappa_\ell^{\mathrm{AS}}$ at 500~K.

\begin{table}
\centering
\caption{Performance metrics for predicting $\log(\kappa_\ell^{\mathrm{TDEP}})$ for the 32 test compounds using the active sampling and baseline models. The metrics are: R2-score, root-mean-square error (RMSE), Spearman correlation, and Pearson correlation.
The standard deviations are in parenthesis.}
\label{tbl:statistics}
\begin{tabular}{lll}
\toprule
 &       Active &     Baseline \\
\midrule
      R2 &  0.84 (0.03) &  0.36 (0.13) \\
    RMSE &  0.21 (0.02) &  0.43 (0.04) \\
      Spearman &  0.85 (0.04) &  0.79 (0.07) \\
      Pearson  &  0.93 (0.02) &  0.64 (0.09) \\
\bottomrule

\end{tabular}
\end{table}

Table~\ref{tbl:statistics} summarizes various performance metrics of the ML models.
The baseline model predictions, $\log(\kappa_\ell^{\mathrm{BL}})$,
has a Spearman correlation of 0.79 with $\log(\kappa_\ell^{\mathrm{TDEP}})$, which is larger than the correlation of Carrete~et~al.~\cite{carreteFindingUnprecedentedlyLowThermalConductivity2014} of 0.74. Their model is based on fewer training samples, but more complex features. 
However, even though the Spearman correlation metric is fair and the low $\kappa_\ell^{\mathrm{TDEP}}$ compounds do tend to be in the lower end of the spectrum of $\log(\kappa_\ell^{\mathrm{BL}})$, the model fails to differentiate between the truly low $\kappa_\ell^{\mathrm{TDEP}}$ compounds and the rest.
The Spearman correlation of the active sampling model increases to 0.85. The superior ability of the active sampling model to predict properties of compounds with low $\kappa_\ell^{\mathrm{TDEP}}$ results in improvement of the other performance metrics as well.

\subsection{\label{sec:active_sample_model_without_DFT_features}Machine learning at tier-0 level}

\begin{figure}[t!]
\includegraphics[width=\linewidth]{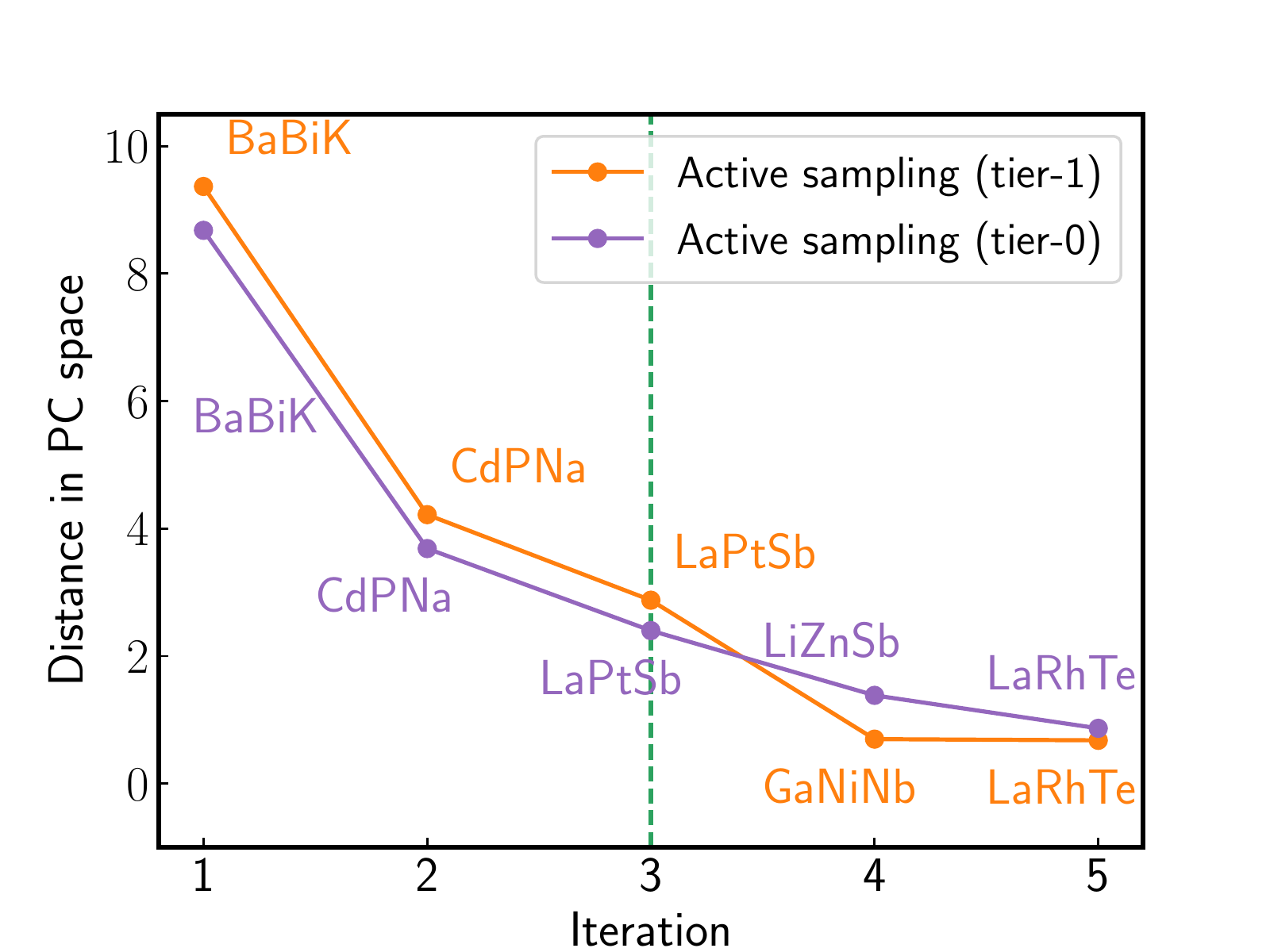}
\caption{\label{fig:PCA_distance}
Distance in PC space between the compound that is farthest from the training pool and its closest training pool neighbor after each iteration of sampling. Orange indicates PC space distance in the space spanned with tier-1 features, and purple shows the results for tier-0 features. The green line indicates the point at which we stop the inclusion of more compounds.}
\end{figure}
\begin{figure}[h!]
\begin{subfigure}[b]{\columnwidth}
\includegraphics[width=71mm, scale=1]{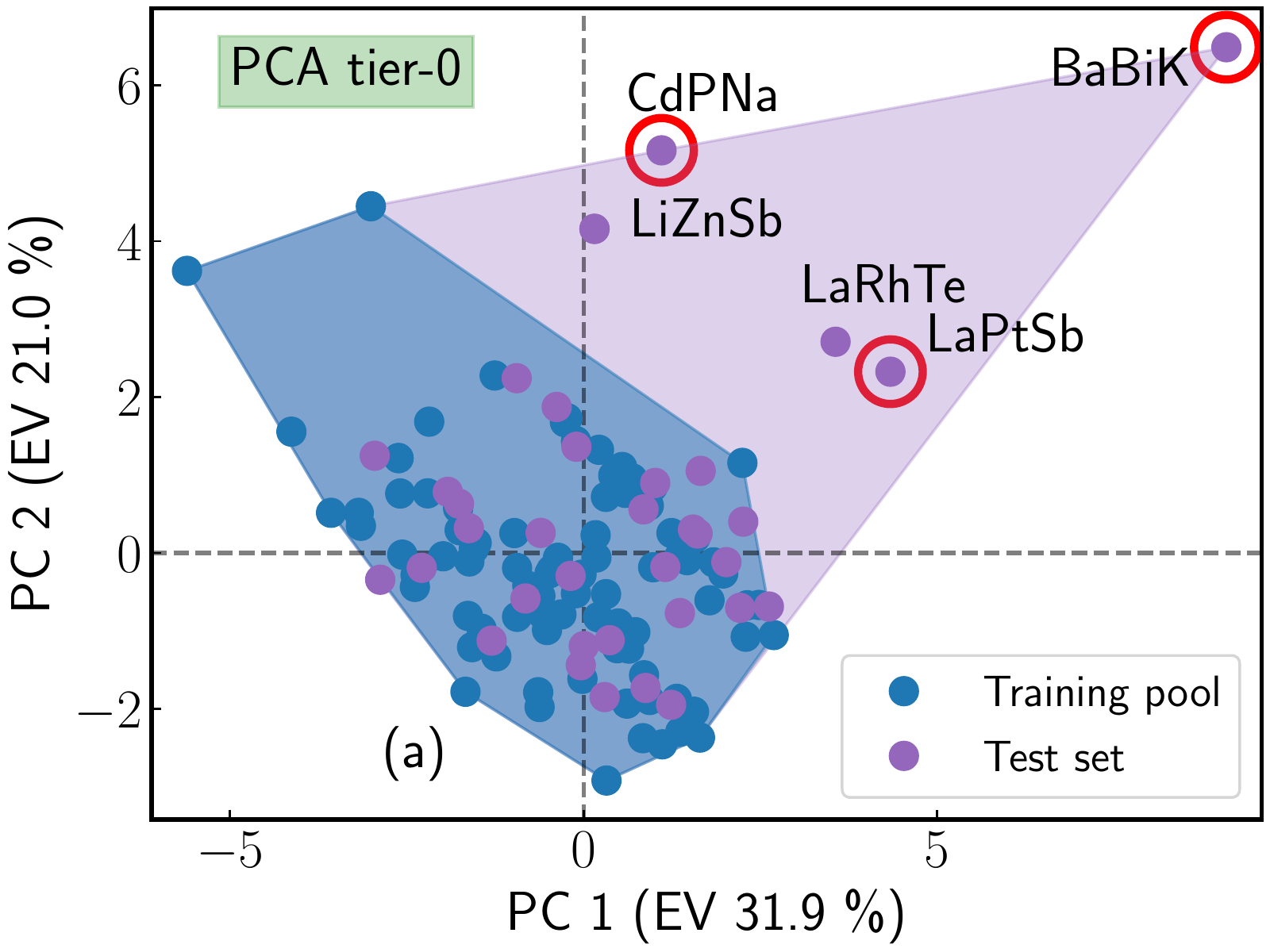}
\caption*{\label{fig:PCA_distance_wo}}
\end{subfigure}
\vspace{-2.25\baselineskip}
\\
\begin{subfigure}[b]{\columnwidth}
\includegraphics[width=71mm, scale=1]{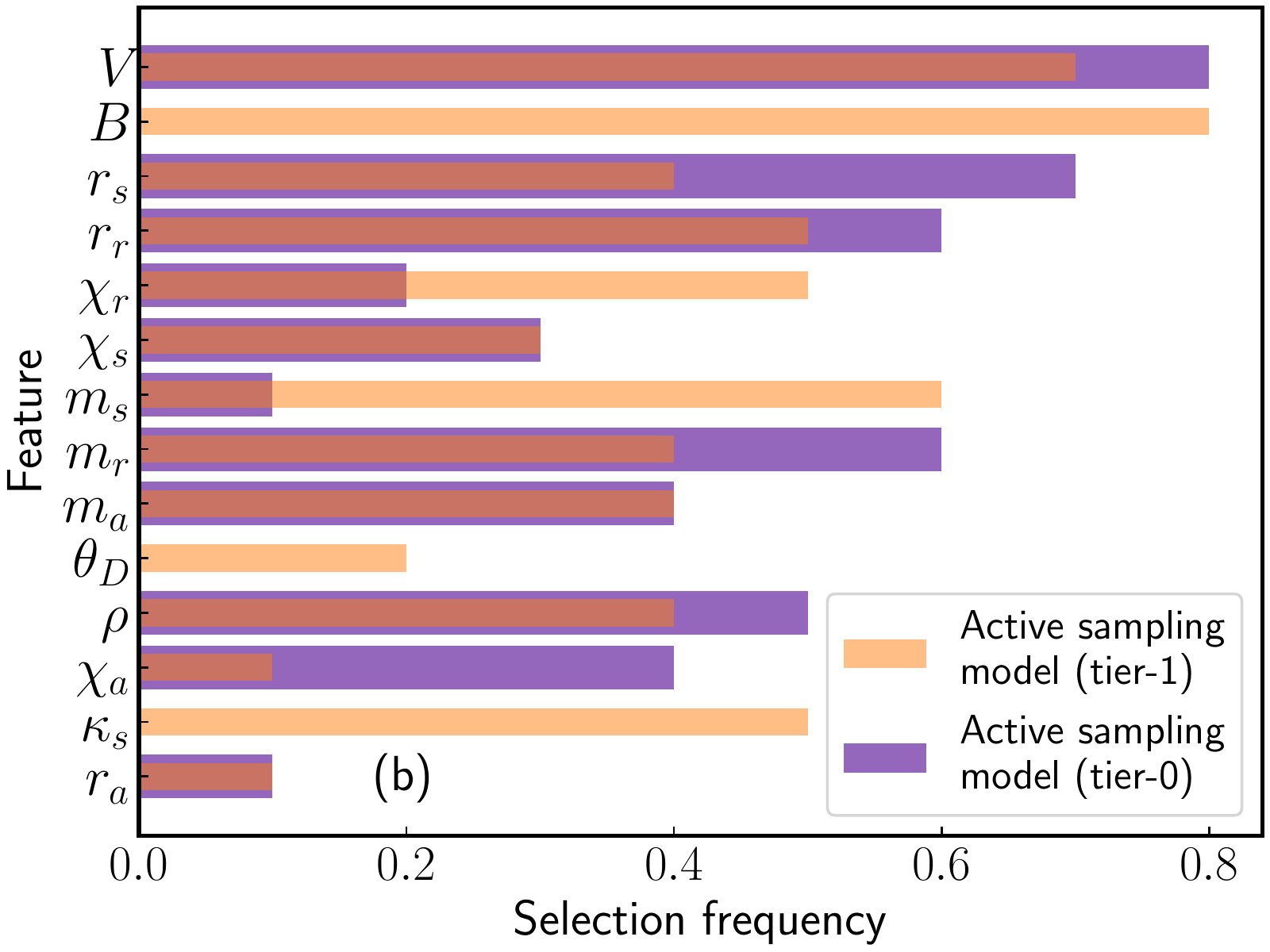}
\caption*{\label{fig:feature_importance_wo}}
\end{subfigure}
\vspace{-2.25\baselineskip}
\\
\begin{subfigure}[b]{\columnwidth}
\includegraphics[width=71mm, scale=1]{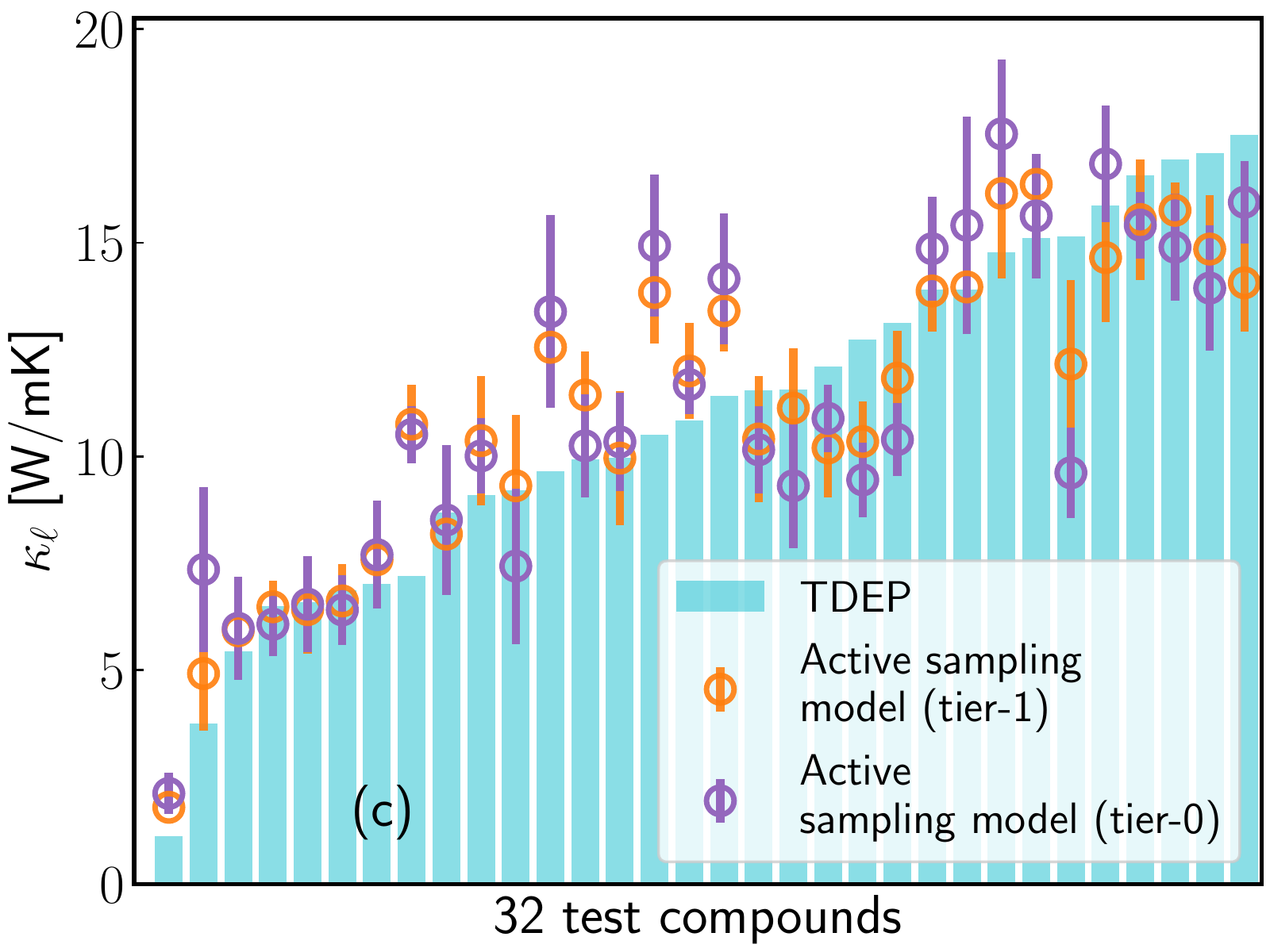}
\caption*{\label{fig:dualplot_wo}}
\end{subfigure}
\vspace{-2.5\baselineskip}
\caption{(a) PCA plot for the compounds using tier-0 features. The horizontal axis shows the first PC while the vertical axis shows the second. (b) Selection frequencies from the EFS using RF regression for the active sampling models using tier-0 and tier-1 features. (c) Predicted and TDEP lattice thermal conductivity using active sampling models based on tier-0 and tier-1 features.}\label{fig:wo_DFT_results}
\end{figure}

The need for DFT-level input can be a drawback of using tier-1 features, as 
in some cases, experimental or calculated lattice constants are known while elastic tensors or bulk moduli are lacking.
Thus, to uncover the potential of ML based on simpler features, we compare compound sampling and ML using tier-0 and tier-1 features. PCA-based active sampling with tier-0 features identifies the same three compounds as found earlier, as shown in Fig.~\ref{fig:PCA_distance} and Fig.~\ref{fig:wo_DFT_results}~(a). 
We allow for LiZnSb to be in the test set in this case, even though the drop-off in PC distance is less steep, as it allows for direct comparison of ML performance when using tier-0 and tier-1 features.  

Fig.~\ref{fig:wo_DFT_results}~(b) compares the EFS selection frequencies of the active sampling models with tier-0 and tier-1 features. The average number of features chosen in the EFS for the active sampling model (tier-0) is 4.7, and increased selection frequencies are seen for $\chi_{a}$ and $r_{s}$. 




Fig.~\ref{fig:wo_DFT_results}~(c) shows that the active sampling model (tier-0) identifies the lowest $\kappa_\ell^{\mathrm{TDEP}}$ compound, but fails to differentiate the second and third lowest from the rest. 
This model has weaker predictivity overall compared to the active sampling model (tier-1), with an R2-score of 0.73 and Spearman correlation of 0.78.
Despite that tier-0 features are used, the active sampling model (tier-0) outperforms the baseline model, underlining the importance of sample selection.

We also note that pre-sampling using PCA with tier-0 features could be used for pruning the test set. This can be done by excluding compounds that lie close to or within already sampled compound clusters in PC space. This would reduce the number of compounds in the test set, and thus reduce the computational resources needed for calculating the tier-1 DFT features.


While the performance gain when using more complex features and larger training sets has been demonstrated in earlier studies~\cite{chenMachineLearningModels2019,junejaCouplingHighThroughputProperty2019,junejaGuidedPatchworkKriging2020,loftisLatticeThermalConductivity2021,wangIdentificationCrystallineMaterials2020,tewariMachineLearningApproaches2020,ZhuCharting2, miyazakiMachineLearningBased2021},
this work demonstrates that using rather modest training set sizes and low feature complexity can give reliable predictions by adopting active sample selection.

On a cautionary note, the use of semi-random selection rather than truly random selection accentuates the performance gains when doing active sampling. We also find that only including one or two of the low lattice thermal conductivity compounds in the models significantly reduces performance compared to including all three. 
Performance with a truly random model would hence be sensitive to exactly which training samples are selected. In any case, a key advantage of PCA is that when used in the process to include additional HH compounds, we have a procedure to identify whether the properties of a given compound can be predicted reliably.

\subsection{\label{sec:comparison_with_experiment}Comparison with experiments}

Computed and predicted lattice thermal conductivities do not always agree perfectly with that of experiments. 
For comparison, the $\kappa_{\ell}^{\mathrm{TDEP}}$ for NbCoSn and ZrNiSn of 12.2~W/mK and 10.8~W/mK are higher than lattice thermal conductivity measured in experiments: 7.0~W/mK~\cite{HeNbCoSn} and 8.7~W/mK~\cite{YanNbCoSn} for NbCoSn; 5.4~W/mK~\cite{BahramiZrNiSn} and 6.1~W/mK~\cite{XieIntrinsicDisorder} for ZrNiSn. Predictions made by the active sampling model (tier-1) for NbCoSn and ZrNiSn are 13.4 W/mK and 12.0 W/mK, respectively, when trained on calculated data. 
At the current level of theory, the difference between the ML predictions and TDEP lattice thermal conductivity is therefore much lower than that of experiment and theory.
Sample dependent phonon scattering mechanisms due to physical properties such as grain boundaries, intrinsic disorder, and antisite defects can drastically reduce lattice thermal conductivity~\cite{SchradeGrainBoundary, XieIntrinsicDisorder, eliassenLatticeThermalConductivity2017, KatreAntisite, CarreteNanograined,HazamaDefect, MiyazakiLocalDistortion}, and can explain why experimentally measured lattice thermal conductivity is lower than predictions when computed lattice thermal conductivity only includes three-phonon and isotope scattering.
Including such scattering mechanisms in the calculated lattice thermal conductivity can give values closer to that of experiment~\cite{eliassenLatticeThermalConductivity2017, GengNanoGrain}, 
which future studies should contemplate including in ML training for making models more representative of the lattice thermal conductivity of real-world samples.

\section{\label{sec:conclusion}Summary and conclusion}
This study has explored strategies for using machine learning for finding low lattice thermal conductivity compounds using a limited number of training samples. Moreover, rather simple features were used, which can be found directly in material databases or computed straightforwardly.
The exploration was made possible by computing 
lattice thermal conductivity with the temperature-dependent effective potential method
for 122 half-Heusler compounds.
We first demonstrated how a model based on a semi-random pool of materials (i.e. assumed "bad luck" in the training set)
was unable to separate the truly low lattice thermal conductivity compounds in the test set from the rest.
To improve the model,
we used active sample selection based on principal component analysis.
This approach suggested three compounds to be included in the training process.
The subsequent inclusions resulted in a substantial improvement of model performance,
in particular the ability to identify 
the remaining low lattice thermal conductivity compounds in the test set. Active sample selection without density functional theory-based features also identified necessary compounds to include in the model, but excluding the features in the model training resulted in weaker predictivity.

Our study demonstrates how active sampling can improve machine learning predictivity by accurately predicting properties of compounds dissimilar from the typical ones in a material class.
More narrowly, we expect the procedure outlined here to be adopted to study broader classes of materials to systematically identify new low lattice thermal conductivity compounds. 



\section{\label{sec:data_availability}Data availability} 

The data forming the basis of this study is available from the authors upon reasonable request. 
 
\section*{\label{sec:acknowldgement}Acknowledgements}
The computations in this work were done on the high performance cluster Saga managed by UNINETT Sigma2. This work is part of the Allotherm project (Project No.  314778) supported by the Research Council of Norway. We further thank Øven Andreas Grimenes for helpful discussion and proofreading. 
 
\appendix
\addtocounter{table}{-1}
\section{\label{app:A}}

Table~\ref{table:kappas} shows $\kappa_\ell^{\mathrm{TDEP}}$ for the HH compounds. Predictions made with the active sampling model (tier-1) are in the parenthesis. 

\begin{table*}[b!]
\centering
\caption{\label{table:kappas}Calculated $\kappa_\ell^{\mathrm{TDEP}}$ at 500~K for the HHs. The 32 compounds in the final test set are highlighted with bold text, and predictions made with the active sampling model (tier-1) are in parenthesis. 
}

\begingroup
\setlength{\tabcolsep}{5pt} 
\renewcommand{\arraystretch}{0.8}
\begin{tabular}{ll | ll | ll | ll}

\toprule
            &       $\kappa_\ell$ [W/mK] &             &         $\kappa_\ell$ [W/mK] &             &         $\kappa_\ell$ [W/mK] &             &         $\kappa_\ell$ [W/mK] \\
\midrule
          LaPtSb &         0.85 &           ZrPtGe &           8.27 &  \textbf{ZrCoBi} &  11.54 (10.41) &           TaCoGe &          15.47 \\
 \textbf{LaRhTe} &  1.11 (1.79) &           TiPdSn &           8.42 &  \textbf{NbIrSn} &  11.57 (11.13) &            VRuAs &          15.52 \\
           BaBiK &         1.99 &  \textbf{NbRuBi} &     8.70 (8.18) &           HfCoAs &          11.64 &           TiPdGe &          15.78 \\
           CdPNa &         2.58 &           TiPtSn &           8.88 &           TiRhSb &          11.72 &  \textbf{GaNiNb} &  15.88 (14.65) \\
 \textbf{LiZnSb} &  3.76 (4.92) &           HfIrAs &           9.04 &           HfRhAs &          11.76 &           NbOsAs &          15.95 \\
          TaIrPb &          5.20 &  \textbf{ZrNiPb} &    9.10 (10.37) &           TaIrSn &          11.84 &           SiCoTa &          16.28 \\
          TiPtPb &         5.43 &            VOsSb &           9.12 &           NbOsSb &          11.86 &   \textbf{VRhGe} &  16.57 (15.54) \\
 \textbf{BiPdSc} &  5.44 (5.92) &           AlSiLi &           9.18 &           ZrRhSb &          11.89 &            VCoGe &          16.58 \\
           BiNiY &         5.46 &  \textbf{ZrPtSn} &     9.20 (9.32) &  \textbf{HfCoBi} &   12.11 (10.20) &           TaRuAs &           16.60 \\
          HfPtPb &         5.48 &           HfPdSn &           9.32 &           NbCoSn &          12.23 &           GaPtTa &          16.75 \\
          HfPdPb &         5.86 &            VRuSb &           9.32 &           TaFeBi &           12.40 &            VOsAs &          16.82 \\
          TaOsBi &          5.90 &           ZrPdSn &           9.37 &           HfCoSb &          12.48 &           TaRhGe &          16.85 \\
          TiPdPb &         5.94 &           AsNiSc &           9.46 &           ZrNiGe &          12.48 &   \textbf{VIrGe} &  16.95 (15.76) \\
          TaRhPb &         5.95 &  \textbf{TaCoPb} &   9.66 (12.55) &           NbRuSb &          12.52 &            GeFeW &          17.07 \\
          NbIrPb &         6.07 &  \textbf{NbRhSn} &   9.92 (11.44) &  \textbf{TiIrSb} &  12.73 (10.35) &  \textbf{TaFeSb} &   17.10 (14.85) \\
 \textbf{ZrPtPb} &   6.50 (6.48) &           NbCoPb &           9.95 &  \textbf{TaRuSb} &  13.12 (11.83) &           AlAuHf &          17.11 \\
          HfIrBi &         6.58 &  \textbf{HfNiPb} &    9.96 (9.96) &           TaCoSn &          13.38 &           NbIrGe &          17.21 \\
 \textbf{ZrPdPb} &  6.59 (6.42) &           HfPtSn &           9.98 &            VFeSb &           13.40 &  \textbf{TiRhAs} &  17.52 (14.06) \\
          TiIrBi &         6.68 &           HfPdGe &          10.28 &           HfIrSb &          13.65 &           TaOsAs &          18.09 \\
 \textbf{ZrIrBi} &  6.88 (6.62) &  \textbf{TiNiSn} &   10.50 (13.83) &           TeFeTi &          13.82 &           TiCoBi &           19.10 \\
          BiNiSc &         6.97 &  \textbf{ZrNiSn} &   10.84 (12.00) &  \textbf{TiCoSb} &   13.90 (13.87) &           NbCoGe &          19.37 \\
          HfRhBi &          7.00 &           ZrIrSb &          10.97 &           TiNiPb &          13.91 &           TaIrGe &          19.57 \\
           VIrSn &         7.01 &           ZrPdGe &          10.98 &  \textbf{ZrCoAs} &  13.91 (13.97) &           TiCoAs &          19.92 \\
 \textbf{TaRuBi} &  7.02 (7.58) &           HfPtGe &          11.08 &           TiIrAs &          14.12 &           NbRuAs &          20.21 \\
           VRhSn &         7.11 &           TeRuZr &          11.11 &           ZrCoSb &          14.29 &           NbCoSi &          20.31 \\
 \textbf{ZrIrAs} &  7.20 (10.76) &           HfNiSn &          11.29 &           TiPtGe &          14.34 &           NbRhGe &          20.63 \\
          NbRhPb &         7.21 &            VCoSn &           11.30 &           ZrRhAs &          14.54 &           NbFeAs &           22.20 \\
          NbOsBi &         7.26 &           TaRhSn &          11.37 &  \textbf{TiNiGe} &  14.78 (16.15) &            VFeAs &           22.60 \\
          ZrRhBi &         7.33 &           NbFeBi &           11.40 &           NbFeSb &           15.00 &            LiBSi &          23.45 \\
          AlGeLi &         7.68 &  \textbf{HfNiGe} &  11.41 (13.41) &  \textbf{TaFeAs} &  15.12 (16.37) &                  &                \\
          TiRhBi &         7.85 &           HfRhSb &          11.49 &  \textbf{TaOsSb} &  15.16 (12.16) &                  &                \\
\bottomrule

\end{tabular}
\endgroup
\end{table*}

\addtocounter{figure}{-6}





\bibliographystyle{elsarticle-num} 
\bibliography{references}

\begin{thebibliography}{10}
\expandafter\ifx\csname url\endcsname\relax
  \def\url#1{\texttt{#1}}\fi
\expandafter\ifx\csname urlprefix\endcsname\relax\def\urlprefix{URL }\fi
\expandafter\ifx\csname href\endcsname\relax
  \def\href#1#2{#2} \def\path#1{#1}\fi

\bibitem{snyderComplexThermoelectricMaterials2008}
G.~J. Snyder, E.~S. Toberer,
  \href{http://www.nature.com/articles/nmat2090}{Complex thermoelectric
  materials}, Nat. Mater. 7~(2) (2008) 105--114.
\newblock \href {https://doi.org/10.1038/nmat2090}
  {\path{doi:10.1038/nmat2090}}.

\bibitem{champierThermoelectricGeneratorsReview2017}
D.~Champier,
  \href{https://linkinghub.elsevier.com/retrieve/pii/S0196890417301851}{Thermoelectric
  generators: {A} review of applications}, Energy Convers. Manag. 140 (2017)
  167--181.
\newblock \href {https://doi.org/10.1016/j.enconman.2017.02.070}
  {\path{doi:10.1016/j.enconman.2017.02.070}}.

\bibitem{morenoReviewRecentProgress2020}
J.~J.~G. Moreno, A review of recent progress in thermoelectric materials
  through computational methods, Mater. Renew. Sustain. Energy (2020) 22\href
  {https://doi.org/10.1007/s40243-020-00175-5}
  {\path{doi:10.1007/s40243-020-00175-5}}.

\bibitem{weiReviewCurrentHighZT2020}
J.~Wei, L.~Yang, Z.~Ma, P.~Song, M.~Zhang, J.~Ma, F.~Yang, X.~Wang,
  \href{http://link.springer.com/10.1007/s10853-020-04949-0}{Review of current
  high-{ZT} thermoelectric materials}, J Mater Sci 55~(27) (2020) 12642--12704.
\newblock \href {https://doi.org/10.1007/s10853-020-04949-0}
  {\path{doi:10.1007/s10853-020-04949-0}}.

\bibitem{recatala-gomezAcceleratedThermoelectricMaterials2020}
J.~Recatala-Gomez, A.~Suwardi, I.~Nandhakumar, A.~Abutaha, K.~Hippalgaonkar,
  \href{https://pubs.acs.org/doi/10.1021/acsaem.9b02222}{Toward {accelerated}
  {thermoelectric} {materials} and {process} {discovery}}, ACS Appl. Energy
  Mater. 3~(3) (2020) 2240--2257.
\newblock \href {https://doi.org/10.1021/acsaem.9b02222}
  {\path{doi:10.1021/acsaem.9b02222}}.

\bibitem{paskovEffectSiDoping2017}
P.~P. Paskov, M.~Slomski, J.~H. Leach, J.~F. Muth, T.~Paskova,
  \href{http://aip.scitation.org/doi/10.1063/1.4989626}{Effect of {Si} doping
  on the thermal conductivity of bulk {GaN} at elevated temperatures – theory
  and experiment}, AIP Advances 7~(9) (2017) 095302.
\newblock \href {https://doi.org/10.1063/1.4989626}
  {\path{doi:10.1063/1.4989626}}.

\bibitem{kimInfluencePdDoping2019}
S.~Y. Kim, H.-S. Kim, K.~H. Lee, H.-j. Cho, S.-s. Choo, S.-w. Hong, Y.~Oh,
  Y.~Yang, K.~Lee, J.-H. Lim, S.-M. Choi, H.~J. Park, W.~H. Shin, S.-i. Kim,
  \href{https://www.mdpi.com/1996-1944/12/24/4080}{Influence of {Pd} {doping}
  on {electrical} and {thermal} {properties} of n-{type}
  $\mathrm{Cu}_{0.008}\mathrm{Bi}_{2}\mathrm{Te}_{2.7}\mathrm{Se}_{0.3}$
  {alloys}}, Materials 12~(24) (2019) 4080.
\newblock \href {https://doi.org/10.3390/ma12244080}
  {\path{doi:10.3390/ma12244080}}.

\bibitem{berlandThermoelectricTransportTrends2019}
K.~Berland, N.~Shulumba, O.~Hellman, C.~Persson, O.~M. Løvvik,
  \href{https://doi.org/10.1063/1.5117288}{Thermoelectric transport trends in
  group 4 half-{H}eusler alloys}, J. Appl. Phys. 126~(14) (2019) 145102.
\newblock \href {http://arxiv.org/abs/https://doi.org/10.1063/1.5117288}
  {\path{arXiv:https://doi.org/10.1063/1.5117288}}.

\bibitem{liHighThroughputScreeningAdvanced2019}
R.~Li, X.~Li, L.~Xi, J.~Yang, D.~J. Singh, W.~Zhang,
  \href{https://pubs.acs.org/doi/10.1021/acsami.9b01196}{High-{throughput}
  {screening} for {advanced} {thermoelectric} {materials}: {diamond}-{like}
  {AB$\mathrm{X}_{2}$} {compounds}}, ACS Appl. Mater. Interfaces 11~(28) (2019)
  24859--24866.
\newblock \href {https://doi.org/10.1021/acsami.9b01196}
  {\path{doi:10.1021/acsami.9b01196}}.

\bibitem{choudharyDatadrivenDiscovery3D2020}
K.~Choudhary, K.~F. Garrity, F.~Tavazza,
  \href{https://iopscience.iop.org/article/10.1088/1361-648X/aba06b}{Data-driven
  discovery of {3D} and {2D} thermoelectric materials}, J. Condens. Matter
  Phys. 32~(47) (2020) 475501.
\newblock \href {https://doi.org/10.1088/1361-648X/aba06b}
  {\path{doi:10.1088/1361-648X/aba06b}}.

\bibitem{liuHighthroughputDescriptorPrediction2020}
J.~Liu, S.~Han, G.~Cao, Z.~Zhou, C.~Sheng, H.~Liu,
  \href{https://iopscience.iop.org/article/10.1088/1361-6463/ab898e}{A
  high-throughput descriptor for prediction of lattice thermal conductivity of
  half-{Heusler} compounds}, J. Phys. D: Appl. Phys. 53~(31) (2020) 315301.
\newblock \href {https://doi.org/10.1088/1361-6463/ab898e}
  {\path{doi:10.1088/1361-6463/ab898e}}.

\bibitem{heUltralowThermalConductivity2016}
J.~He, M.~Amsler, Y.~Xia, S.~S. Naghavi, V.~I. Hegde, S.~Hao, S.~Goedecker,
  V.~Ozoliņš, C.~Wolverton,
  \href{https://link.aps.org/doi/10.1103/PhysRevLett.117.046602}{Ultralow
  {thermal} {conductivity} in {full} {Heusler} {semiconductors}}, Phys. Rev.
  Lett. 117~(4) (2016) 046602.
\newblock \href {https://doi.org/10.1103/PhysRevLett.117.046602}
  {\path{doi:10.1103/PhysRevLett.117.046602}}.

\bibitem{raghuvanshiHighThroughputSearch2020}
P.~R. Raghuvanshi, S.~Mondal, A.~Bhattacharya,
  \href{http://xlink.rsc.org/?DOI=D0TA06810A}{A high throughput search for
  efficient thermoelectric half-{Heusler} compounds}, J. Mater. Chem. A 8~(47)
  (2020) 25187--25197.
\newblock \href {https://doi.org/10.1039/D0TA06810A}
  {\path{doi:10.1039/D0TA06810A}}.

\bibitem{JiaScreeningPromisingThermoelectric}
T.-T. Jia, Z.~Feng, S.~Guo, X.~Zhang, Y.~Zhang, Screening promising
  thermoelectric materials in binary chalcogenides through high-throughput
  computations, ACS Appl. Mater. Interfaces 12 (2020) 11852--11864.
\newblock \href {https://doi.org/10.1021/acsami.9b23297}
  {\path{doi:10.1021/acsami.9b23297}}.

\bibitem{togoDistributionsPhononLifetimes2015}
A.~Togo, L.~Chaput, I.~Tanaka,
  \href{https://link.aps.org/doi/10.1103/PhysRevB.91.094306}{Distributions of
  phonon lifetimes in {Brillouin} zones}, Phys. Rev. B 91~(9) (2015) 094306.
\newblock \href {https://doi.org/10.1103/PhysRevB.91.094306}
  {\path{doi:10.1103/PhysRevB.91.094306}}.

\bibitem{hellmanTemperaturedependentEffectiveThirdorder2013}
O.~Hellman, I.~A. Abrikosov,
  \href{https://link.aps.org/doi/10.1103/PhysRevB.88.144301}{Temperature-dependent
  effective third-order interatomic force constants from first principles},
  Phys. Rev. B 88~(14) (2013) 144301.
\newblock \href {https://doi.org/10.1103/PhysRevB.88.144301}
  {\path{doi:10.1103/PhysRevB.88.144301}}.

\bibitem{zhouLatticeAnharmonicityThermal2014}
F.~Zhou, W.~Nielson, Y.~Xia, V.~Ozoliņš,
  \href{https://link.aps.org/doi/10.1103/PhysRevLett.113.185501}{Lattice
  {anharmonicity} and {thermal} {conductivity} from {compressive} {sensing} of
  {first}-{principles} {calculations}}, Phys. Rev. Lett. 113~(18) (2014)
  185501.
\newblock \href {https://doi.org/10.1103/PhysRevLett.113.185501}
  {\path{doi:10.1103/PhysRevLett.113.185501}}.

\bibitem{carreteFindingUnprecedentedlyLowThermalConductivity2014}
J.~Carrete, W.~Li, N.~Mingo, S.~Wang, S.~Curtarolo,
  \href{https://link.aps.org/doi/10.1103/PhysRevX.4.011019}{Finding
  {unprecedentedly} {low}-{thermal}-{conductivity} {half}-{Heusler}
  {semiconductors} via {high}-{throughput} {materials} {modeling}}, Phys. Rev.
  X 4~(1) (2014) 011019.
\newblock \href {https://doi.org/10.1103/PhysRevX.4.011019}
  {\path{doi:10.1103/PhysRevX.4.011019}}.

\bibitem{juExploringDiamondlikeLattice2019}
S.~Ju, R.~Yoshida, C.~Liu, K.~Hongo, T.~Tadano, Exploring diamond-like lattice
  thermal conductivity crystals via feature-based transfer learning (2019)
  27\href {https://doi.org/10.1103/PhysRevMaterials.5.053801}
  {\path{doi:10.1103/PhysRevMaterials.5.053801}}.

\bibitem{chenMachineLearningModels2019}
L.~Chen, H.~Tran, R.~Batra, C.~Kim, R.~Ramprasad,
  \href{https://linkinghub.elsevier.com/retrieve/pii/S0927025619304549}{Machine
  learning models for the lattice thermal conductivity prediction of inorganic
  materials}, Comput. Mater. Sci. 170 (2019) 109155.
\newblock \href {https://doi.org/10.1016/j.commatsci.2019.109155}
  {\path{doi:10.1016/j.commatsci.2019.109155}}.

\bibitem{wangIdentificationCrystallineMaterials2020}
X.~Wang, S.~Zeng, Z.~Wang, J.~Ni,
  \href{https://pubs.acs.org/doi/abs/10.1021/acs.jpcc.9b11610}{Identification
  of {crystalline} {materials} with {ultra}-{low} {thermal} {conductivity}
  {based} on {machine} {learning} {study}}, J. Phys. Chem. C 124~(16) (2020)
  8488--8495.
\newblock \href {https://doi.org/10.1021/acs.jpcc.9b11610}
  {\path{doi:10.1021/acs.jpcc.9b11610}}.

\bibitem{loftisLatticeThermalConductivity2021}
C.~Loftis, K.~Yuan, Y.~Zhao, M.~Hu, J.~Hu,
  \href{https://pubs.acs.org/doi/10.1021/acs.jpca.0c08103}{Lattice {thermal}
  {conductivity} {prediction} {using} {symbolic} {regression} and {machine}
  {learning}}, J. Phys. Chem. A 125~(1) (2021) 435--450.
\newblock \href {https://doi.org/10.1021/acs.jpca.0c08103}
  {\path{doi:10.1021/acs.jpca.0c08103}}.

\bibitem{ZhuCharting2}
T.~Zhu, R.~He, S.~Gong, T.~Xie, P.~Gorai, K.~Nielsch, J.~C. Grossman, Charting
  lattice thermal conductivity for inorganic crystals and discovering rare
  earth chalcogenides for thermoelectrics, Energy Environ. Sci. 14 (2021)
  3559--3566.
\newblock \href {https://doi.org/10.1039/D1EE00442E}
  {\path{doi:10.1039/D1EE00442E}}.

\bibitem{visariaMachinelearningassistedSpacetransformationAccelerates2020}
D.~Visaria, A.~Jain,
  \href{http://aip.scitation.org/doi/10.1063/5.0028241}{Machine-learning-assisted
  space-transformation accelerates discovery of high thermal conductivity
  alloys}, Appl. Phys. Lett. 117~(20) (2020) 202107.
\newblock \href {https://doi.org/10.1063/5.0028241}
  {\path{doi:10.1063/5.0028241}}.

\bibitem{Gaultois2}
C.~M. Collins, L.~M. Daniels, Q.~Gibson, M.~W. Gaultois, M.~Moran, R.~Feetham,
  M.~J. Pitcher, M.~S. Dyer, C.~Delacotte, M.~Zanella, C.~A. Murray, G.~Glodan,
  O.~Pérez, D.~Pelloquin, T.~D. Manning, J.~Alaria, G.~R. Darling, J.~B.
  Claridge, M.~J. Rosseinsky, Discovery of a low thermal conductivity oxide
  guided by probe structure prediction and machine learning, Angew. Chem. Int.
  Ed (2021).
\newblock \href {https://doi.org/https://doi.org/10.1002/anie.202102073}
  {\path{doi:https://doi.org/10.1002/anie.202102073}}.

\bibitem{gaultoisRecommendationEngineSuggesting2016}
M.~W. Gaultois, A.~O. Oliynyk, A.~Mar, T.~D. Sparks, G.~J. Mulholland,
  B.~Meredig, \href{http://arxiv.org/abs/1502.07635}{A recommendation engine
  for suggesting unexpected thermoelectric chemistries}, APL Mater. 4~(5)
  (2016) 053213, arXiv: 1502.07635.
\newblock \href {https://doi.org/10.1063/1.4952607}
  {\path{doi:10.1063/1.4952607}}.

\bibitem{casperHalfHeuslerCompoundsNovel2012}
F.~Casper, T.~Graf, S.~Chadov, B.~Balke, C.~Felser,
  \href{https://iopscience.iop.org/article/10.1088/0268-1242/27/6/063001}{Half-{Heusler}
  compounds: novel materials for energy and spintronic applications}, Semicond.
  Sci. Technol. 27~(6) (2012) 063001.
\newblock \href {https://doi.org/10.1088/0268-1242/27/6/063001}
  {\path{doi:10.1088/0268-1242/27/6/063001}}.

\bibitem{bosHalfHeuslerThermoelectricsComplex2014}
J.-W.~G. Bos, R.~A. Downie,
  \href{https://iopscience.iop.org/article/10.1088/0953-8984/26/43/433201}{Half-{Heusler}
  thermoelectrics: a complex class of materials}, J. Condens. Matter Phys.
  26~(43) (2014) 433201.
\newblock \href {https://doi.org/10.1088/0953-8984/26/43/433201}
  {\path{doi:10.1088/0953-8984/26/43/433201}}.

\bibitem{zhuHighEfficiencyHalfHeusler2015}
T.~Zhu, C.~Fu, H.~Xie, Y.~Liu, X.~Zhao,
  \href{http://doi.wiley.com/10.1002/aenm.201500588}{High {efficiency}
  {half}-{Heusler} {thermoelectric} {materials} for {energy} {harvesting}},
  Adv. Energy Mater. 5~(19) (2015) 1500588.
\newblock \href {https://doi.org/10.1002/aenm.201500588}
  {\path{doi:10.1002/aenm.201500588}}.

\bibitem{yuanEffectsSbSubstitution2017}
B.~Yuan, B.~Wang, L.~Huang, X.~Lei, L.~Zhao, C.~Wang, Q.~Zhang,
  \href{http://link.springer.com/10.1007/s11664-016-5168-z}{Effects of {Sb}
  {substitution} by {Sn} on the {thermoelectric} {properties} of {ZrCoSb}}, J.
  Electron. Mater. 46~(5) (2017) 3076--3082.
\newblock \href {https://doi.org/10.1007/s11664-016-5168-z}
  {\path{doi:10.1007/s11664-016-5168-z}}.

\bibitem{horiFirstprinciplesCalculationLattice2020}
A.~Hori, S.~Minami, M.~Saito, F.~Ishii,
  \href{http://aip.scitation.org/doi/10.1063/1.5143038}{First-principles
  calculation of lattice thermal conductivity and thermoelectric figure of
  merit in ferromagnetic half-{Heusler} alloy {CoMnSb}}, Appl. Phys. Lett.
  116~(24) (2020) 242408.
\newblock \href {https://doi.org/10.1063/1.5143038}
  {\path{doi:10.1063/1.5143038}}.

\bibitem{chauhanDefectEngineeringEnhancement2020}
N.~S. Chauhan, P.~R. Raghuvanshi, K.~Tyagi, K.~K. Johari, L.~Tyagi, B.~Gahtori,
  S.~Bathula, A.~Bhattacharya, S.~D. Mahanti, V.~N. Singh, Y.~V. Kolen’ko,
  A.~Dhar, \href{https://pubs.acs.org/doi/abs/10.1021/acs.jpcc.0c00681}{Defect
  {engineering} for {enhancement} of {thermoelectric} {performance} of ({Zr},
  {Hf}){NiSn}-{Based} n-type {half}-{Heusler} {alloys}}, J. Phys. Chem. C
  124~(16) (2020) 8584--8593.
\newblock \href {https://doi.org/10.1021/acs.jpcc.0c00681}
  {\path{doi:10.1021/acs.jpcc.0c00681}}.

\bibitem{zhuDiscoveryTaFeSbbasedHalfHeuslers2019}
H.~Zhu, J.~Mao, Y.~Li, J.~Sun, Y.~Wang, Q.~Zhu, G.~Li, Q.~Song, J.~Zhou, Y.~Fu,
  R.~He, T.~Tong, Z.~Liu, W.~Ren, L.~You, Z.~Wang, J.~Luo, A.~Sotnikov, J.~Bao,
  K.~Nielsch, G.~Chen, D.~J. Singh, Z.~Ren,
  \href{http://www.nature.com/articles/s41467-018-08223-5}{Discovery of
  {TaFeSb}-based half-{Heuslers} with high thermoelectric performance}, Nat.
  Commun. 10~(1) (2019) 270.
\newblock \href {https://doi.org/10.1038/s41467-018-08223-5}
  {\path{doi:10.1038/s41467-018-08223-5}}.

\bibitem{sunRemarkablyHighThermoelectric2020}
H.-L. Sun, C.-L. Yang, M.-S. Wang, X.-G. Ma,
  \href{https://pubs.acs.org/doi/10.1021/acsami.9b19198}{Remarkably {high}
  {thermoelectric} {efficiencies} of the {half}-{Heusler} {compounds} {BXGa}
  ({X} = {Be}, {Mg}, and {Ca})}, ACS Appl. Mater. Interfaces 12~(5) (2020)
  5838--5846.
\newblock \href {https://doi.org/10.1021/acsami.9b19198}
  {\path{doi:10.1021/acsami.9b19198}}.

\bibitem{zhaoSynthesisThermoelectricProperties2014}
D.~Zhao, M.~Zuo, Z.~Wang, X.~Teng, H.~Geng,
  \href{https://www.worldscientific.com/doi/abs/10.1142/S1793604714500325}{Synthesis
  and thermoelectric properties of tantalum-doped {ZrNiSn} half-{Heusler}
  alloys}, Funct. Mater. Lett. 07~(03) (2014) 1450032.
\newblock \href {https://doi.org/10.1142/S1793604714500325}
  {\path{doi:10.1142/S1793604714500325}}.

\bibitem{Zhou2018LargeTP}
J.~Zhou, H.~Zhu, T.-H. Liu, Q.~Song, R.~He, J.~Mao, Z.~Liu, W.~Ren, B.~Liao,
  D.~J. Singh, Z.~Ren, G.~Chen, Large thermoelectric power factor from crystal
  symmetry-protected non-bonding orbital in half-heuslers, Nat. Commun. 9
  (2018).
\newblock \href {https://doi.org/10.1038/s41467-018-03866-w}
  {\path{doi:10.1038/s41467-018-03866-w}}.

\bibitem{fengCharacterizationRattlingRelation2020}
Z.~Feng, Y.~Fu, Y.~Zhang, D.~J. Singh,
  \href{http://arxiv.org/abs/2001.08029}{Characterization of rattling in
  relation to thermal conductivity: ordered half-{Heusler} semiconductors},
  Phys. Rev. B 101~(6) (2020) 064301, arXiv: 2001.08029.
\newblock \href {https://doi.org/10.1103/PhysRevB.101.064301}
  {\path{doi:10.1103/PhysRevB.101.064301}}.

\bibitem{xueLaPtSbHalfHeuslerCompound2016}
Q.~Y. Xue, H.~J. Liu, D.~D. Fan, L.~Cheng, B.~Y. Zhao, J.~Shi,
  \href{http://xlink.rsc.org/?DOI=C6CP03211G}{{LaPtSb}: a half-{Heusler}
  compound with high thermoelectric performance}, Phys. Chem. Chem. Phys.
  18~(27) (2016) 17912--17916.
\newblock \href {https://doi.org/10.1039/C6CP03211G}
  {\path{doi:10.1039/C6CP03211G}}.

\bibitem{hanHighThermoelectricPerformance2020}
S.~H. Han, Z.~Z. Zhou, C.~Y. Sheng, J.~H. Liu, L.~Wang, H.~M. Yuan, H.~J. Liu,
  \href{https://iopscience.iop.org/article/10.1088/1361-648X/aba2e7}{High
  thermoelectric performance of half-{Heusler} compound {BiBaK} with
  intrinsically low lattice thermal conductivity}, J. Condens. Matter Phys.
  32~(42) (2020) 425704.
\newblock \href {https://doi.org/10.1088/1361-648X/aba2e7}
  {\path{doi:10.1088/1361-648X/aba2e7}}.

\bibitem{samsonidzeAcceleratedScreeningThermoelectric2018}
G.~Samsonidze, B.~Kozinsky,
  \href{http://doi.wiley.com/10.1002/aenm.201800246}{Accelerated {screening} of
  {thermoelectric} {materials} by {first}-{principles} {computations} of
  {electron}-{phonon} {scattering}}, Adv. Energy Mater. 8~(20) (2018) 1800246.
\newblock \href {https://doi.org/10.1002/aenm.201800246}
  {\path{doi:10.1002/aenm.201800246}}.

\bibitem{zhangFirstprinciplesStudyLayered2019}
S.~Zhang, B.~Xu, Y.~Lin, C.~Nan, W.~Liu,
  \href{http://xlink.rsc.org/?DOI=C9RA00247B}{First-principles study of the
  layered thermoelectric material {TiNBr}}, RSC Adv. 9~(23) (2019)
  12886--12894.
\newblock \href {https://doi.org/10.1039/C9RA00247B}
  {\path{doi:10.1039/C9RA00247B}}.

\bibitem{sarkarFerroelectricInstabilityInduced2020}
D.~Sarkar, T.~Ghosh, S.~Roychowdhury, R.~Arora, S.~Sajan, G.~Sheet, U.~V.
  Waghmare, K.~Biswas,
  \href{https://pubs.acs.org/doi/10.1021/jacs.0c03696}{Ferroelectric
  {instability} {induced} {ultralow} {thermal} {conductivity} and {high}
  {thermoelectric} {performance} in {rhombohedral} \textit{p} -{Type} {GeSe}
  {crystal}}, J. Am. Chem. Soc. 142~(28) (2020) 12237--12244.
\newblock \href {https://doi.org/10.1021/jacs.0c03696}
  {\path{doi:10.1021/jacs.0c03696}}.

\bibitem{zhaoUltralowThermalConductivity2014}
L.-D. Zhao, S.-H. Lo, Y.~Zhang, H.~Sun, G.~Tan, C.~Uher, C.~Wolverton, V.~P.
  Dravid, M.~G. Kanatzidis,
  \href{http://www.nature.com/articles/nature13184}{Ultralow thermal
  conductivity and high thermoelectric figure of merit in {SnSe} crystals},
  Nature 508~(7496) (2014) 373--377.
\newblock \href {https://doi.org/10.1038/nature13184}
  {\path{doi:10.1038/nature13184}}.

\bibitem{kresseInitioMolecularDynamics1993}
G.~Kresse, J.~Hafner,
  \href{https://link.aps.org/doi/10.1103/PhysRevB.47.558}{\textit{{Ab} initio}
  molecular dynamics for liquid metals}, Phys. Rev. B 47~(1) (1993) 558--561.
\newblock \href {https://doi.org/10.1103/PhysRevB.47.558}
  {\path{doi:10.1103/PhysRevB.47.558}}.

\bibitem{kresseEfficiencyAbinitioTotal1996}
G.~Kresse, J.~Furthmüller,
  \href{https://linkinghub.elsevier.com/retrieve/pii/0927025696000080}{Efficiency
  of ab-initio total energy calculations for metals and semiconductors using a
  plane-wave basis set}, Comput. Mater. Sci. 6~(1) (1996) 15--50.
\newblock \href {https://doi.org/10.1016/0927-0256(96)00008-0}
  {\path{doi:10.1016/0927-0256(96)00008-0}}.

\bibitem{kresseEfficientIterativeSchemes1996}
G.~Kresse, J.~Furthmüller,
  \href{https://link.aps.org/doi/10.1103/PhysRevB.54.11169}{Efficient iterative
  schemes for \textit{ab initio} total-energy calculations using a plane-wave
  basis set}, Phys. Rev. B 54~(16) (1996) 11169--11186.
\newblock \href {https://doi.org/10.1103/PhysRevB.54.11169}
  {\path{doi:10.1103/PhysRevB.54.11169}}.

\bibitem{perdewRestoringDensityGradientExpansion2008}
J.~P. Perdew, A.~Ruzsinszky, G.~I. Csonka, O.~A. Vydrov, G.~E. Scuseria, L.~A.
  Constantin, X.~Zhou, K.~Burke,
  \href{https://link.aps.org/doi/10.1103/PhysRevLett.100.136406}{Restoring the
  {density}-{gradient} {expansion} for {exchange} in {solids} and {surfaces}},
  Phys. Rev. Lett. 100~(13) (2008) 136406.
\newblock \href {https://doi.org/10.1103/PhysRevLett.100.136406}
  {\path{doi:10.1103/PhysRevLett.100.136406}}.

\bibitem{csonkaAssessingPerformanceRecent2009}
G.~I. Csonka, J.~P. Perdew, A.~Ruzsinszky, P.~H.~T. Philipsen, S.~Lebègue,
  J.~Paier, O.~A. Vydrov, J.~G. Ángyán,
  \href{https://link.aps.org/doi/10.1103/PhysRevB.79.155107}{Assessing the
  performance of recent density functionals for bulk solids}, Phys. Rev. B
  79~(15) (2009) 155107.
\newblock \href {https://doi.org/10.1103/PhysRevB.79.155107}
  {\path{doi:10.1103/PhysRevB.79.155107}}.

\bibitem{hellmanLatticeDynamicsAnharmonic2011}
O.~Hellman, I.~A. Abrikosov, S.~I. Simak,
  \href{https://link.aps.org/doi/10.1103/PhysRevB.84.180301}{Lattice dynamics
  of anharmonic solids from first principles}, Phys. Rev. B 84~(18) (2011)
  180301.
\newblock \href {https://doi.org/10.1103/PhysRevB.84.180301}
  {\path{doi:10.1103/PhysRevB.84.180301}}.

\bibitem{tamuraIsotopeScatteringDispersive1983}
S.-i. Tamura, \href{https://link.aps.org/doi/10.1103/PhysRevB.27.858}{Isotope
  scattering of dispersive phonons in {Ge}}, Phys. Rev. B 27~(2) (1983)
  858--866.
\newblock \href {https://doi.org/10.1103/PhysRevB.27.858}
  {\path{doi:10.1103/PhysRevB.27.858}}.

\bibitem{tamuraIsotopeScatteringLargewavevector1984}
S.-i. Tamura, \href{https://link.aps.org/doi/10.1103/PhysRevB.30.849}{Isotope
  scattering of large-wave-vector phonons in {GaAs} and {InSb}:
  {Deformation}-dipole and overlap-shell models}, Phys. Rev. B 30~(2) (1984)
  849--854.
\newblock \href {https://doi.org/10.1103/PhysRevB.30.849}
  {\path{doi:10.1103/PhysRevB.30.849}}.

\bibitem{shulumbaIntrinsicLocalized}
N.~Shulumba, O.~Hellman, A.~J. Minnich,
  \href{https://link.aps.org/doi/10.1103/PhysRevB.95.014302}{Intrinsic
  localized mode and low thermal conductivity of {PbSe}}, Phys. Rev. B 95
  (2017) 014302.
\newblock \href {https://doi.org/10.1103/PhysRevB.95.014302}
  {\path{doi:10.1103/PhysRevB.95.014302}}.

\bibitem{AndersonSimplifiedMethod}
O.~{Anderson}, {A simplified method for calculating the debye temperature from
  elastic constants}, J Phys Chem Solids 24~(7) (1963) 909--917.
\newblock \href {https://doi.org/10.1016/0022-3697(63)90067-2}
  {\path{doi:10.1016/0022-3697(63)90067-2}}.

\bibitem{breiman_rf}
L.~Breiman, \href{https://doi.org/10.1023/A:1010933404324}{Random forests},
  Mach. Learn. 45~(1) (2001) 5–32.
\newblock \href {https://doi.org/10.1023/A:1010933404324}
  {\path{doi:10.1023/A:1010933404324}}.

\bibitem{python_ml}
S.~Raschka, V.~Mirjalili, Python machine learning: machine learning and deep
  learning with Python, Scikit-Learn, and TensorFlow, 2nd Edition, 2nd Edition,
  Packt Publishing, 2017.

\bibitem{jabbarMethodsAvoidOverFitting2014}
H.~K. Jabbar, R.~Z. Khan,
  \href{http://rpsonline.com.sg/proceedings/9789810952471/html/017}{Methods to
  avoid over-fitting and under-fitting in supervised machine learning
  (comparative study)}, in: Computer {Science}, {Communication} and
  {Instrumentation} {Devices}, Research Publishing Services, 2014, pp.
  163--172.
\newblock \href {https://doi.org/10.3850/978-981-09-5247-1\_017}
  {\path{doi:10.3850/978-981-09-5247-1\_017}}.

\bibitem{raschkaMLxtendProvidingMachine2018}
S.~Raschka, \href{http://joss.theoj.org/papers/10.21105/joss.00638}{{MLxtend}:
  {Providing} machine learning and data science utilities and extensions to
  {Python}’s scientific computing stack}, JOSS 3~(24) (2018) 638.
\newblock \href {https://doi.org/10.21105/joss.00638}
  {\path{doi:10.21105/joss.00638}}.

\bibitem{pedregosaScikitlearnMachineLearning2011}
F.~Pedregosa, G.~Varoquaux, A.~Gramfort, V.~Michel, B.~Thirion, O.~Grisel,
  M.~Blondel, P.~Prettenhofer, R.~Weiss, V.~Dubourg, J.~Vanderplas, A.~Passos,
  D.~Cournapeau, Scikit-learn: {machine} {learning} in {Python}, J. Mach.
  Learn. Res. 12 (2011) 2825.

\bibitem{tomicHoggormPythonLibrary2019}
O.~Tomic, T.~Graff, K.~Liland, T.~Næs,
  \href{http://joss.theoj.org/papers/10.21105/joss.00980}{Hoggorm: a python
  library for explorative multivariate statistics}, JOSS 4~(39) (2019) 980.
\newblock \href {https://doi.org/10.21105/joss.00980}
  {\path{doi:10.21105/joss.00980}}.

\bibitem{miyazakiMachineLearningBased2021}
H.~Miyazaki, T.~Tamura, M.~Mikami, K.~Watanabe, N.~Ide, O.~M. Ozkendir,
  N.~Toichi, Machine learning based prediction of lattice thermal conductivity
  for half-{Heusler} compounds using atomic information, Sci. Rep. (2021).
\newblock \href {https://doi.org/10.1038/s41598-021-92030-4}
  {\path{doi:10.1038/s41598-021-92030-4}}.

\bibitem{LiangCRYSPNet}
H.~Liang, V.~Stanev, A.~G. Kusne, I.~Takeuchi, {CRYSPNet}: Crystal structure
  predictions via neural networks, Phys. Rev. Mater. 4 (2020) 123802.
\newblock \href {https://doi.org/10.1103/PhysRevMaterials.4.123802}
  {\path{doi:10.1103/PhysRevMaterials.4.123802}}.

\bibitem{LiMlatticeabc}
Y.~Li, W.~Yang, R.~Dong, J.~Hu, {MLatticeABC}: Generic lattice constant
  prediction of crystal materials using machine learning, ACS Omega 6 (2021)
  11585 -- 11594.
\newblock \href {https://doi.org/10.1021/acsomega.1c00781}
  {\path{doi:10.1021/acsomega.1c00781}}.

\bibitem{MeiJaAtomicWeights}
J.~Meija, T.~Coplen, M.~Berglund, W.~Brand, P.~De~Bièvre, M.~Gröning,
  N.~Holden, J.~Irrgeher, R.~Loss, T.~Walczyk, T.~Prohaska, Atomic weights of
  the elements 2013 (iupac technical report), Pure Appl. Chem. 88 (2016)
  265--291.
\newblock \href {https://doi.org/10.1515/pac-2015-0305}
  {\path{doi:10.1515/pac-2015-0305}}.

\bibitem{AllenElectronegativity}
L.~C. Allen, Electronegativity is the average one-electron energy of the
  valence-shell electrons in ground-state free atoms, J. Am. Chem. Soc. 111
  (1989) 9003--9014.
\newblock \href {https://doi.org/10.1021/JA00207A003}
  {\path{doi:10.1021/JA00207A003}}.

\bibitem{CorderoCovalentRadii}
B.~Cordero, V.~Gómez, A.~Platero-Prats, M.~Revés, J.~Echeverria, E.~Cremades,
  F.~Barragán, S.~Alvarez, Covalent radii revisited, Dalton trans. 21 (2008)
  2832--8.
\newblock \href {https://doi.org/10.1039/b801115j}
  {\path{doi:10.1039/b801115j}}.

\bibitem{slackNonmetallicRystalsHigh1973}
G.~A. Slack, Nonmetallic crystals with high thermal conductivity, J Phys Chem
  Solids 34~(2) (1973) 15.
\newblock \href {https://doi.org/10.1016/0022-3697(73)90092-9}
  {\path{doi:10.1016/0022-3697(73)90092-9}}.

\bibitem{jiaLatticeThermalConductivity2017}
T.~Jia, G.~Chen, Y.~Zhang,
  \href{http://link.aps.org/doi/10.1103/PhysRevB.95.155206}{Lattice thermal
  conductivity evaluated using elastic properties}, Phys. Rev. B 95~(15) (2017)
  155206.
\newblock \href {https://doi.org/10.1103/PhysRevB.95.155206}
  {\path{doi:10.1103/PhysRevB.95.155206}}.

\bibitem{juMaterialsInformaticsHeat2019}
S.~Ju, J.~Shiomi,
  \href{https://www.tandfonline.com/doi/full/10.1080/15567265.2019.1576816}{Materials
  {informatics} for {heat} {transfer}: {recent} {progresses} and
  {perspectives}}, Nanoscale Microscale Thermophys. Eng. 23~(2) (2019)
  157--172.
\newblock \href {https://doi.org/10.1080/15567265.2019.1576816}
  {\path{doi:10.1080/15567265.2019.1576816}}.

\bibitem{jainCommentaryMaterialsProject2013}
A.~Jain, S.~P. Ong, G.~Hautier, W.~Chen, W.~D. Richards, S.~Dacek, S.~Cholia,
  D.~Gunter, D.~Skinner, G.~Ceder, K.~A. Persson,
  \href{http://aip.scitation.org/doi/10.1063/1.4812323}{Commentary: {The}
  {Materials} {Project}: {A} materials genome approach to accelerating
  materials innovation}, APL Mater. 1~(1) (2013) 011002.
\newblock \href {https://doi.org/10.1063/1.4812323}
  {\path{doi:10.1063/1.4812323}}.

\bibitem{gautierPredictionAcceleratedLaboratory2015}
R.~Gautier, X.~Zhang, L.~Hu, L.~Yu, Y.~Lin, T.~O.~L. Sunde, D.~Chon, K.~R.
  Poeppelmeier, A.~Zunger,
  \href{http://www.nature.com/articles/nchem.2207}{Prediction and accelerated
  laboratory discovery of previously unknown 18-electron {ABX} compounds},
  Nature Chem 7~(4) (2015) 308--316.
\newblock \href {https://doi.org/10.1038/nchem.2207}
  {\path{doi:10.1038/nchem.2207}}.

\bibitem{gaultois3}
M.~W. Gaultois, T.~D. Sparks, How much improvement in thermoelectric
  performance can come from reducing thermal conductivity?, Appl. Phys. Lett.
  104~(11) (2014) 113906.
\newblock \href {https://doi.org/10.1063/1.4869232}
  {\path{doi:10.1063/1.4869232}}.

\bibitem{anInitioPhononDispersions2008}
J.~An, A.~Subedi, D.~Singh,
  \href{https://linkinghub.elsevier.com/retrieve/pii/S0038109808005486}{Ab
  initio phonon dispersions for {PbTe}}, Solid State Commun. 148~(9-10) (2008)
  417--419.
\newblock \href {https://doi.org/10.1016/j.ssc.2008.09.027}
  {\path{doi:10.1016/j.ssc.2008.09.027}}.

\bibitem{junejaCouplingHighThroughputProperty2019}
R.~Juneja, G.~Yumnam, S.~Satsangi, A.~K. Singh,
  \href{https://pubs.acs.org/doi/10.1021/acs.chemmater.9b01046}{Coupling the
  {high}-{throughput} {property} {map} to {machine} {learning} for {predicting}
  {lattice} {thermal} {conductivity}}, Chem. Mater. 31~(14) (2019) 5145--5151.
\newblock \href {https://doi.org/10.1021/acs.chemmater.9b01046}
  {\path{doi:10.1021/acs.chemmater.9b01046}}.

\bibitem{junejaGuidedPatchworkKriging2020}
R.~Juneja, A.~K. Singh,
  \href{https://iopscience.iop.org/article/10.1088/2515-7639/ab78f2}{Guided
  patchwork kriging to develop highly transferable thermal conductivity
  prediction models}, J. Phys. Mater. 3~(2) (2020) 024006.
\newblock \href {https://doi.org/10.1088/2515-7639/ab78f2}
  {\path{doi:10.1088/2515-7639/ab78f2}}.

\bibitem{tewariMachineLearningApproaches2020}
A.~Tewari, S.~Dixit, N.~Sahni, S.~P. Bordas,
  \href{https://www.cambridge.org/core/product/identifier/S2632673620000076/type/journal\_article}{Machine
  learning approaches to identify and design low thermal conductivity oxides
  for thermoelectric applications}, DCE 1 (2020) e8.
\newblock \href {https://doi.org/10.1017/dce.2020.7}
  {\path{doi:10.1017/dce.2020.7}}.

\bibitem{HeNbCoSn}
R.~He, L.~Huang, Y.~Wang, G.~Samsonidze, B.~Kozinsky, Q.~Zhang, Z.~Ren,
  Enhanced thermoelectric properties of n-type {N}b{C}o{S}n half-{H}eusler by
  improving phase purity, APL Mater. 4 (2016) 104804.
\newblock \href {https://doi.org/10.1063/1.4952994}
  {\path{doi:10.1063/1.4952994}}.

\bibitem{YanNbCoSn}
R.~Yan, W.~Xie, B.~Balke, G.~Chen, A.~Weidenkaff, Realizing p-type {N}b{C}o{S}n
  half-{H}eusler compounds with enhanced thermoelectric performance via {S}c
  substitution, Sci. Technol. Adv. Mater. 21 (2020).
\newblock \href {https://doi.org/10.1080/14686996.2020.1726715}
  {\path{doi:10.1080/14686996.2020.1726715}}.

\bibitem{BahramiZrNiSn}
A.~Bahrami, P.~Ying, U.~Wolff, N.~P. Rodríguez, G.~Schierning, K.~Nielsch,
  R.~He, Reduced lattice thermal conductivity for half-{H}eusler {ZrNiSn}
  through cryogenic mechanical alloying, ACS Appl. Mater. Interfaces 13~(32)
  (2021) 38561--38568.
\newblock \href {https://doi.org/10.1021/acsami.1c05639}
  {\path{doi:10.1021/acsami.1c05639}}.

\bibitem{XieIntrinsicDisorder}
H.~Xie, H.~Wang, C.~Fu, Y.~Liu, G.~Snyder, X.~Zhao, T.~Zhu, The intrinsic
  disorder related alloy scattering in {ZrNiSn} half-{H}eusler thermoelectric
  materials, Sci. Rep. 4 (2014) 6888.
\newblock \href {https://doi.org/10.1038/srep06888}
  {\path{doi:10.1038/srep06888}}.

\bibitem{SchradeGrainBoundary}
M.~Schrade, K.~Berland, S.~Eliassen, M.~Guzik, C.~Echevarria-Bonet, M.~Sørby,
  P.~Jenus, B.~Hauback, R.~Tofan, A.~Gunnæs, C.~Persson, O.~Løvvik,
  T.~Finstad, The role of grain boundary scattering in reducing the thermal
  conductivity of polycrystalline {XNiSn} ({X} = {Hf}, {Zr}, {Ti})
  half-{H}eusler alloys, Sci. Rep. 7 (2017).
\newblock \href {https://doi.org/10.1038/s41598-017-14013-8}
  {\path{doi:10.1038/s41598-017-14013-8}}.

\bibitem{eliassenLatticeThermalConductivity2017}
S.~N.~H. Eliassen, A.~Katre, G.~K.~H. Madsen, C.~Persson, O.~M. L\o{}vvik,
  K.~Berland,
  \href{https://link.aps.org/doi/10.1103/PhysRevB.95.045202}{Lattice thermal
  conductivity of
  $\mathrm{Ti}_{x}\mathrm{Zr}_{y}\mathrm{Hf}_{1\ensuremath{-}x\ensuremath{-}y}\mathrm{NiSn}$
  half-heusler alloys calculated from first principles: Key role of nature of
  phonon modes}, Phys. Rev. B 95 (2017) 045202.
\newblock \href {https://doi.org/10.1103/PhysRevB.95.045202}
  {\path{doi:10.1103/PhysRevB.95.045202}}.

\bibitem{KatreAntisite}
A.~Katre, J.~Carrete, N.~Mingo, Unraveling the dominant phonon scattering
  mechanism in thermoelectric compound {ZrNiSn}, J. Mater. Chem. A 4 (07 2016).
\newblock \href {https://doi.org/10.1039/C6TA05868J}
  {\path{doi:10.1039/C6TA05868J}}.

\bibitem{CarreteNanograined}
J.~Carrete~Montaña, N.~Mingo, S.~Wang, S.~Curtarolo, Nanograined
  half-{H}eusler semiconductors as advanced thermoelectrics: An ab initio
  high-throughput statistical study, Adv. Funct. Mater. 24 (2014).
\newblock \href {https://doi.org/10.1002/adfm.201401201}
  {\path{doi:10.1002/adfm.201401201}}.

\bibitem{HazamaDefect}
H.~Hazama, M.~Matsubara, R.~Asahi, T.~Takeuchi, Improvement of thermoelectric
  properties for half-{H}eusler {TiNiSn} by interstitial {Ni} defects, J. Appl.
  Phys. 110 (2011).
\newblock \href {https://doi.org/10.1063/1.3633518}
  {\path{doi:10.1063/1.3633518}}.

\bibitem{MiyazakiLocalDistortion}
H.~Miyazaki, O.~M. Ozkendir, S.~Günaydın, K.~Watanabe, K.~Soda, Y.~Nishino,
  Probing local distortion around structural defects in half-{H}eusler
  thermoelectric {NiZrSn} alloy, Sci. Rep. 10 (2020) 19820.
\newblock \href {https://doi.org/10.1038/s41598-020-76554-9}
  {\path{doi:10.1038/s41598-020-76554-9}}.

\bibitem{GengNanoGrain}
H.~Geng, X.~Meng, H.~Zhang, J.~Zhang, Lattice thermal conductivity of
  nanograined half-{H}eusler solid solutions, Appl. Phys. Lett. 104~(20) (2014)
  202104.
\newblock \href {https://doi.org/10.1063/1.4879248}
  {\path{doi:10.1063/1.4879248}}.

\end{thebibliography}





\end{document}